\documentclass[11pt]{article}
\usepackage{jheppub}
\usepackage{amsmath,amssymb,amsfonts,graphicx,slashed,color, mathtools, amsthm}
\usepackage{comment}
\usepackage{enumerate}
\usepackage{float}
\usepackage{caption}
\usepackage{subcaption}

\newcommand{\be}{\begin{equation}}
\newcommand{\ee}{\end{equation}}
\newcommand{\bea}{\begin{eqnarray}}
\newcommand{\eea}{\end{eqnarray}}
\newcommand{\beas}{\begin{eqnarray*}}
\newcommand{\eeas}{\end{eqnarray*}}
\newcommand{\ba}{\begin{array}}
\newcommand{\ea}{\end{array}}

\title{Bottom-up holographic models for cosmology}

\author[]{Chris Waddell}

\affiliation[]{Department of Physics and Astronomy, University of British Columbia,\\
6224 Agricultural Road, Vancouver, B.C.\ V6T 1Z1, Canada.}

\emailAdd{cwaddell@phas.ubc.ca}

\abstract{In this note, we investigate some simple generalizations of a bottom-up holographic approach to cosmology introduced in arXiv:1810.10601. Our models utilize the Karch/Randall/Takayanagi ansatz for the gravitational dual of a boundary conformal field theory, involving pure AdS gravity and an end-of-the-world brane. Following a suggestion made in arXiv:2102.05057, we consider models with an additional interface brane in the bulk. 
We find that solutions with a viable cosmological interpretation exist only if our model is further generalized, for example by including an Einstein-Hilbert term in the ETW brane action. The physical validity of such models is discussed from the perspective of the effective theory. }


\begin{document}

\maketitle

\parskip=10pt

\newpage

\section{Introduction}

An important open question in theoretical physics is how to formulate a non-perturbative quantum mechanical description of gravity in cosmological backgrounds. Given the theoretical successes of the AdS/CFT correspondence over the past two decades \cite{Maldacena:1997re}, an especially appealing prospect is the possibility of embedding cosmological physics in AdS/CFT, though the viability of this approach for ``realistic" cosmologies remains unclear at present. A number of differing holographic approaches to cosmology appear in the literature; an incomplete catalogue of these includes \cite{Strominger:2001pn, Banks:2001px, Hertog:2004rz, Alishahiha:2004md, Freivogel:2005qh, McFadden:2009fg}.

\begin{figure}
    \centering
    \includegraphics[height=8cm]{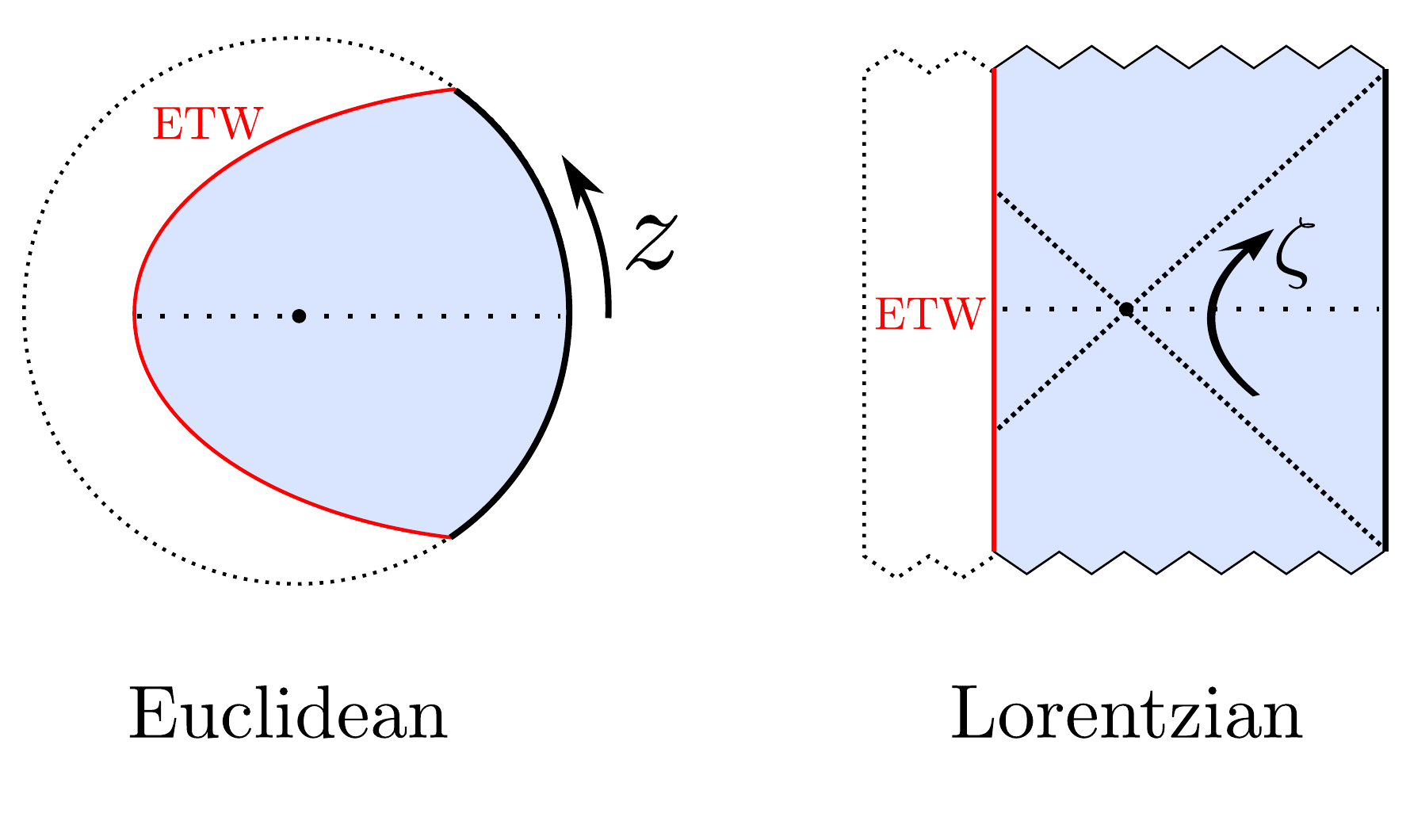}
    \caption{An approach to holographic cosmology proposed in \cite{Cooper:2018cmb}. We begin on the left with a Euclidean BCFT path integral (bold black line), with some choice of boundary condition imposed in the past and future Euclidean time. The transverse directions suppressed in this figure could be taken to have $S^{d}$ or $\mathbb{R}^{d}$ symmetry, so that the Euclidean CFT path integral is on a cylinder or a strip respectively. Cutting open this path integral at the moment of time symmetry, we obtain some state $|\Psi \rangle$ of the holographic CFT. 
    In the bulk, we have a Euclidean asymptotically AdS spacetime (blue) terminating on an ETW brane (red). We may then analytically continue to Lorentzian time to obtain the leading geometry encoding the evolution of $|\Psi \rangle$, shown on the right. The ETW brane stays behind the horizon of an AdS black hole; it is a ``big bang/big crunch" cosmology (with spherical or flat spatial sections). 
   The construction is time-symmetric throughout, with the moment of time symmetry illustrated as a dotted line. Here, $z$ indicates the Euclidean coordinate analytically continued to the Lorentzian time $\zeta$.} 
   \label{fig:BHMC}
\end{figure}

The class of holographic models that we will be interested in here originated with \cite{Cooper:2018cmb}, and has subsequently been further studied in \cite{Antonini:2019qkt, VanRaamsdonk:2020tlr,VanRaamsdonk:2021qgv}. 
In the model considered in these papers, a Euclidean boundary conformal field theory (BCFT) path integral is used to prepare a state of a holographic CFT; via a simple effective or ``bottom-up" model for AdS/BCFT introduced in \cite{Karch:2000gx, Takayanagi:2011zk, Fujita:2011fp}, this state is understood to correspond to an AdS black hole terminating on an end-of-the-world (ETW) brane behind the horizon.
The worldvolume of this ETW brane is a recollapsing (negative cosmological constant) FRW universe. 
Under appropriate conditions, when the ETW brane propagates far outside the black hole horizon in the second asymptotic region, the effective theory on the ETW brane would be expected to exhibit gravity localization via the Karch/Randall/Sundrum mechanism \cite{Randall:1999vf, Karch:2000ct}; the upshot is that gravitational physics on a cosmological background is encoded in a particular state, prepared by a Euclidean path integral, in a holographic theory. See Figure \ref{fig:BHMC} for a visualization of this logic; references \cite{Cooper:2018cmb, VanRaamsdonk:2020tlr, VanRaamsdonk:2021qgv} should be consulted for additional details. 

\begin{figure}
    \centering
    \includegraphics[height=7cm]{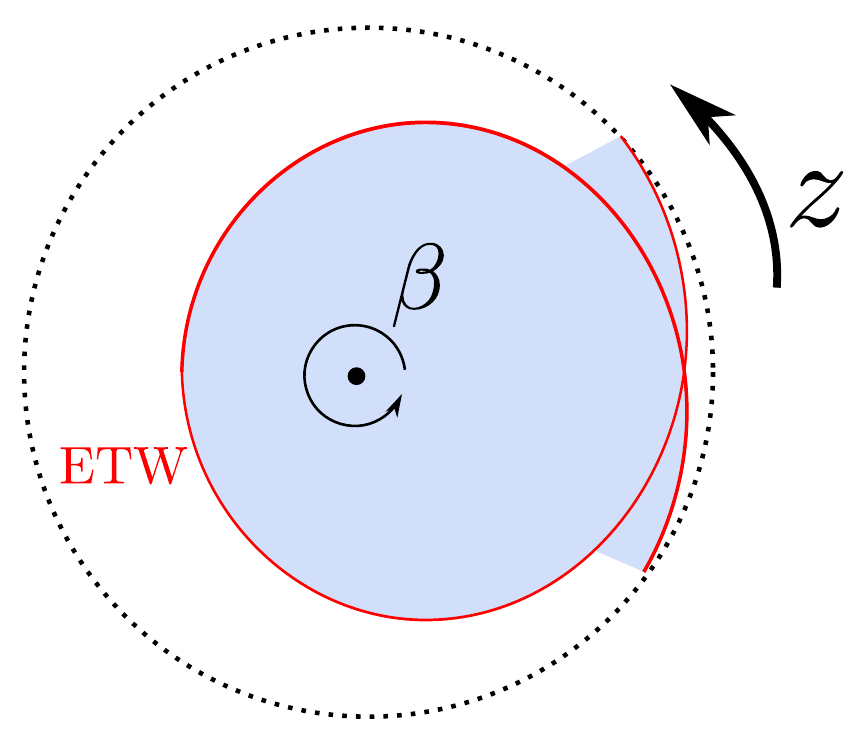}
    \caption{Pathological Euclidean gravity solution with a self-intersecting ETW brane (red). The trajectory of the ETW brane in the Euclidean asymptotically AdS spacetime (blue) can be determined from the equations of motion; the fact that this trajectory self-intersects arises from the coordinate periodicity $z \sim z + \beta$ which must be imposed to ensure smoothness at the coordinate horizon (central dot). } 
   \label{fig:selfint}
\end{figure}

The simple model analyzed in the references mentioned above has proven interesting and suggestive, but not entirely satisfactory: the properties required for the solution to exhibit gravity localization cannot actually be realized within the parameter space.\footnote{The exception to this point is \cite{Antonini:2019qkt}, in which it was found that an ETW brane propagating in a charged black hole background could enjoy the desired properties for cosmology. It is not clear how to make sense of this set-up as an analytic continuation of Euclidean AdS/CFT, since it appears that the gauge field component $A^{0}$ should be imaginary in the Euclidean signature solution.} 
In particular, analytically continuing the Lorentzian solutions where gravity localization is expected to Euclidean signature, we find that the corresponding Euclidean solutions involve self-intersecting ETW branes, whose holographic interpretation is not clear; see Figure \ref{fig:selfint}.

\begin{figure}
    \centering
    \includegraphics[height=5cm]{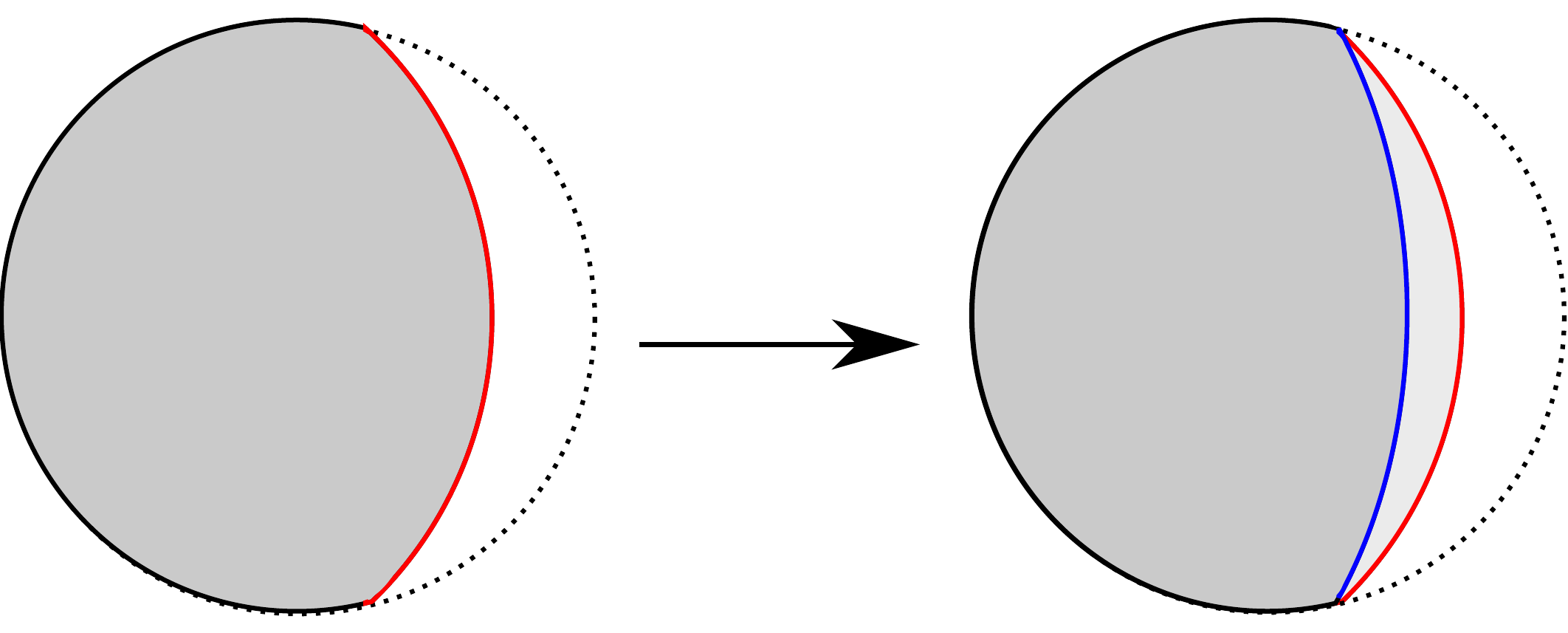}
    \caption{Two putative bulk duals of holographic BCFT. 
    Here, ETW branes are shown in red, and interface branes in blue; the shaded region is an asymptotically AdS Euclidean spacetime. 
    The premise of this work is to move from the model depicted on the left to that depicted on the right, i.e. to introduce an additional interface brane. }
    \label{fig:addinterface}
\end{figure}

An approach to circumventing this issue was proposed by Van Raamsdonk in \cite{VanRaamsdonk:2021qgv}. It was suggested that the previous bottom-up models could be modified by adding an additional ``interface brane" separating two regions of asymptotically AdS spacetime in the bulk, generally with differing AdS lengths $L_{1}$ and $L_{2}$, as shown in Figure \ref{fig:addinterface}. A practical rationale for this proposition is to avoid the self-intersection problem mentioned above, which arises because the Euclidean gravity solutions require a periodically identified coordinate $z \sim z + \beta$ to avoid developing a singularity at the coordinate horizon; in the case with both an ETW brane and an interface brane, the region between these branes no longer includes a coordinate horizon, and therefore need not have any periodically identified coordinate. 

\begin{figure}
    \centering
    \includegraphics[height=6cm]{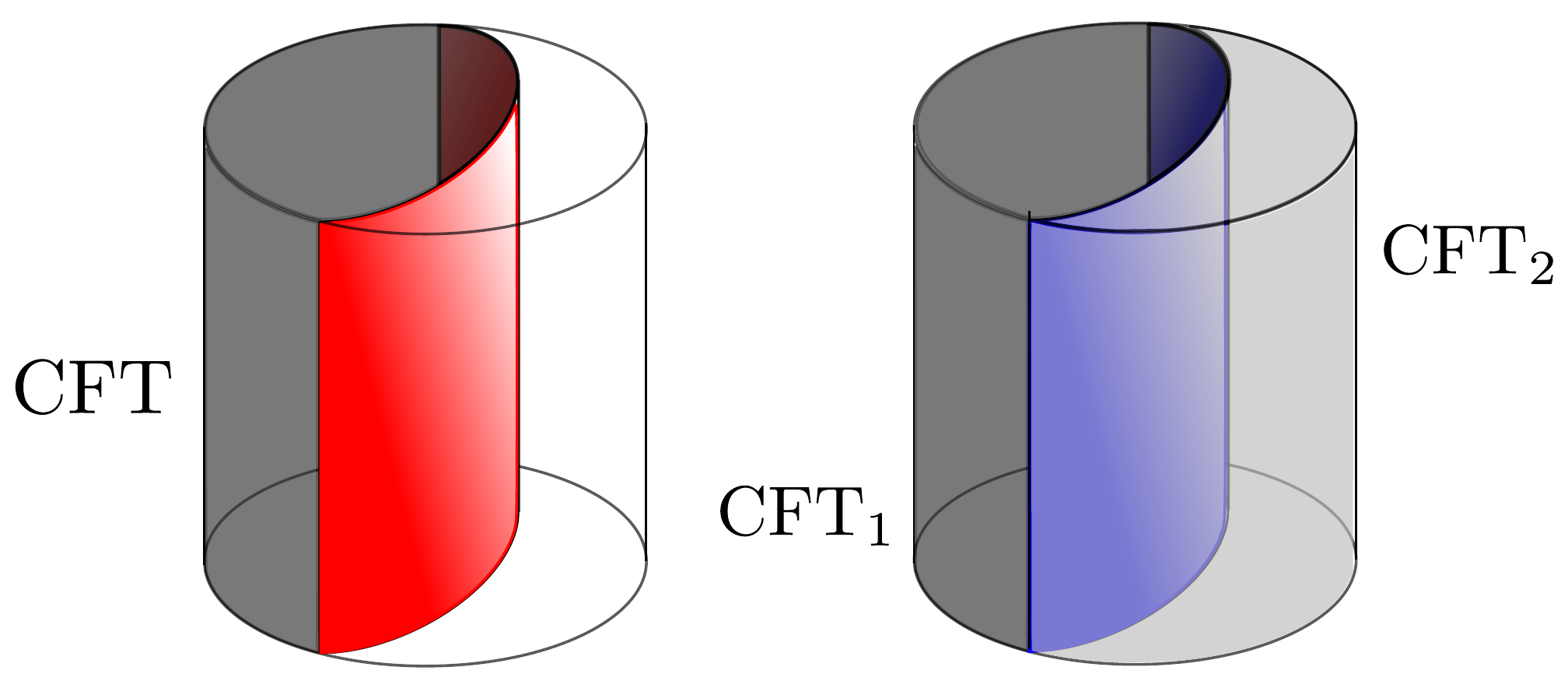}
    \caption{Holographic duals of (left) boundary CFT and (right) interface CFT. The ETW brane is illustrated in red, and the interface brane in blue. We can interpret these diagrams as either representing Euclidean spacetimes, or the Lorentzian spacetimes obtained by Wick rotating a coordinate of one of the transverse directions suppressed in Figures \ref{fig:BHMC}, \ref{fig:selfint}, and \ref{fig:addinterface}, which is the vertical direction here. In Lorentzian signature, the intrinsic geometry of the ETW/interface brane is a traversable asymptotically AdS wormhole.} 
   \label{fig:sols}
\end{figure}

A somewhat more sophisticated motivation was also given in \cite{VanRaamsdonk:2021qgv}, making use of an effect observed in \cite{May:2021xhz}.
To understand the second motivation, one should note that, by performing a different analytic continuation of the bulk Euclidean solutions with a single ETW brane, corresponding to Wick rotating one of the transverse coordinates suppressed in Figures \ref{fig:BHMC}, \ref{fig:selfint}, and \ref{fig:addinterface} (which we assume to have $\mathbb{R}^{d-1}$ planar symmetry for a $(d+1)$-dimensional bulk), one obtains a static Lorentzian solution with an ETW brane whose worldvolume is an asymptotically AdS traversable wormhole; see Figure \ref{fig:sols}. Consequently, the effective description of the cosmology is related by ``double analytic continuation" to an effective theory involving a cutoff CFT on a traversable wormhole background; from this perspective, the non-existence of the solutions relevant for cosmology appears to be related to a no-go result for such traversable wormholes in the absence of large amounts of negative energy \cite{Freivogel:2019lej}. 
However, in a simple bottom-up model for the holographic dual of a conformal interface between two CFTs (also shown in Figure \ref{fig:sols}),
%involving such an interface brane, 
the authors of \cite{May:2021xhz} found that
one could produce an anomalously large negative Casimir energy in one of the two CFTs in a particular critical limit of the tension of a bulk interface brane. From this interface CFT starting point, the model that we are concerned with in this paper would correspond to ``coupling one of the CFTs to gravity" by introducing an ETW ``Planck brane" in the bulk. 
In this case, one might hope that a similar ``negative energy enhancement" effect could allow for a means of negating the hypotheses of the aforementioned no-go result. 

The purpose of this work is to investigate this possibility, generalizing the model of \cite{Cooper:2018cmb} by adding an interface brane. 
%see Figure \ref{fig:addinterface}. 
We begin by considering the case where this interface brane is governed by a single tension parameter; in this case, we argue that there are no consistent solutions in the region of parameter space where we expect to recover gravity localization in the cosmology, suggesting that this model has no significant advantage over the previous model. In particular, putative solutions do not have an ETW brane and an interface brane which join properly; for example, they may instead intersect. We then generalize the model further by incorporating Einstein-Hilbert terms on the ETW brane,\footnote{This is referred to as a ``DGP term" in \cite{Chen:2020uac}, after an analogous construction by Dvali, Gabadadze and Porrati \cite{Dvali:2000hr}, though of course the present model has an asymptotically AdS bulk.} arguing that solutions with the desirable properties should exist in this case. We comment on the nature of the relevant region of parameter space from the perspective of physics in the effective theory on the ETW brane, but leave further commentary about the physicality of this region, and an exploration of the parameter space more broadly, to future work.

The outline of this paper is as follows. In Section \ref{sec:review}, we attempt to briefly review the relevant results already appearing in the literature. We follow this in Section \ref{sec:constantt} with an analysis of the model with an additional interface brane of constant tension, and then further augment this model in Section \ref{sec:EH} with an Einstein-Hilbert term on the ETW brane. We briefly conclude in Section \ref{sec:conclusion}. 

\textbf{Note:} As this work was nearing completion, we were alerted to the existence of similar work by Seamus Fallows and Simon Ross \cite{Fallows:2022ioc}. These authors have graciously agreed to coordinate in submitting pre-prints.

\section{Review of bottom-up holographic solutions for boundary/interface CFT} \label{sec:review}

To keep our presentation self-contained, we will review the relevant holographic models and solutions in this section, and briefly recapitulate some important results in this and the following section. The models discussed in this section follow a prescription for AdS/BCFT involving ETW/interface branes which originated in \cite{Karch:2000gx, Takayanagi:2011zk, Fujita:2011fp}, and the solutions we discuss in this section appear in \cite{Cooper:2018cmb, Simidzija:2020ukv, VanRaamsdonk:2021qgv, May:2021xhz}; the purpose of this section is to summarize the pertinent information from the latter references, and to establish notation. The gravity solutions discussed in Section \ref{sec:BHMC} and \ref{sec:int} correspond to those in the first and second panels of Figure \ref{fig:sols} respectively: they are Euclidean asymptotically AdS$_{d+1}$ spacetimes, with either an ETW brane or an interface brane, and preserving a transverse $\mathbb{R}^{d-1}$ symmetry.

\subsection{Solutions with an ETW brane} \label{sec:BHMC}

We begin by considering a class of models for the gravitational dual of a holographic BCFT, determined by the Euclidean gravitational action
\begin{equation} \label{eq:action_bound}
    \begin{split}
        S & = S_{\textnormal{bulk}} + S_{\textnormal{ETW}}^{\textnormal{matter}} \\
        S_{\textnormal{bulk}} & = \frac{1}{16 \pi G_{\textnormal{bulk}}} \int_{\mathcal{M}} d^{d+1} x \sqrt{g} \: \left( R - 2 \Lambda \right) +  \frac{1}{8 \pi G_{\textnormal{bulk}}} \int_{\textnormal{ETW}} d^{d} y \sqrt{h} \: K \: ,
    \end{split}
\end{equation}
where we take the brane matter action to be
\begin{equation}
    S_{\textnormal{ETW}}^{\textnormal{matter}} = \frac{(1-d) \lambda}{8 \pi G_{\textnormal{bulk}}} \int_{\textnormal{ETW}} d^{d} y \sqrt{ h } \: .
\end{equation}
The cosmological constant $\Lambda$ is related to the AdS length $L$ by
\begin{equation} \label{eq:cosmoconstant}
    \Lambda = - \frac{d (d-1)}{2 L^{2}} \: .
\end{equation}
Here and throughout, we will take $\lambda$ to lie in the interval $\left( 0, \frac{1}{L} \right)$.

The bulk equation of motion is simply the Einstein equation with cosmological constant $\Lambda$; meanwhile, the ETW brane trajectory is given by the equation of motion (see Appendix \ref{app:branetrajectories})
\begin{equation}
    K_{ab} = \lambda h_{ab} \: .
\end{equation}

In \cite{Cooper:2018cmb}, Euclidean solutions with a $S^{d-1}$ spherical symmetry were considered; here, we will instead consider Euclidean solutions with a $\mathbb{R}^{d-1}$ symmetry, though the two cases are completely analogous. The appropriate bulk ansatz is then the \textit{Euclidean AdS soliton} solution
\begin{equation}
    ds^{2} = L^{2} f(r) dz^{2} + \frac{dr^{2}}{f(r)} + r^{2} dx_{\mu} dx^{\mu} \: , \qquad f(r) = \frac{r^{2}}{L^{2}} - \frac{\mu}{r^{d-2}} \: . 
\end{equation} 
The radial coordinate $r$ ranges from the coordinate horizon value $r_{\textnormal{H}} = (\mu L^{2})^{1/d}$ to infinity. In order to avoid a conical singulariy, the $z$ coordinate must be taken to be periodic, with period\footnote{In the solutions of interest to us here, this coordinate horizon is kept in our solution, rather than being excised by the ETW brane, so this periodicity must be enforced.}
\begin{equation} \label{eq:zperiod}
    \beta = \frac{4 \pi L}{d r_{\textnormal{H}}} \: .
\end{equation}

The ETW brane has trajectory $z = z^{\textnormal{ETW}}(r)$ in this (Euclidean) background, determined by the equation of motion (see Appendix \ref{app:branetrajectories})
\begin{equation} \label{eq:ETW_traj}
    \left( \frac{dz^{\textnormal{ETW}}}{dr} \right)^{2} = \frac{\lambda^{2} r^{2}}{L^{2} f(r)^{2}} \frac{1}{f(r) - \lambda^{2} r^{2}} \: .
\end{equation}
In particular, the ETW brane attains a minimum radius at $r_{0}^{\textnormal{ETW}}$ with
\begin{equation} \label{eq:r0ETW}
    f(r_{0}^{\textnormal{ETW}}) = \lambda^{2} (r_{0}^{\textnormal{ETW}})^{2} \: , \qquad r_{0}^{\textnormal{ETW}} = \frac{r_{\textnormal{H}}}{\left( 1 - \lambda^{2} L^{2} \right)^{1/d}} \: .
\end{equation}
We will also denote the $z$-coordinate distance traversed by the ETW brane from its minimum radius to infinity by
\begin{equation} \label{eq:deltaz_ETW}
    \Delta z^{\textnormal{ETW}} \equiv \int_{r_{0}^{\textnormal{ETW}}}^{\infty} dr \: \frac{dz^{\textnormal{ETW}}}{dr} \: .
\end{equation}

Despite the appearance that $r_{0}^{\textnormal{ETW}}$ can be made arbitrarily large by sending $\lambda \rightarrow L^{-1}$, one must recall that the $z$ coordinate is periodic, and such solutions have the ETW brane self-intersecting at finite $r$ in the case $d>2$,\footnote{For $d=2$, the ETW brane always spans coordinate range $2 \Delta z^{\textnormal{ETW}} = \frac{\beta}{2}$, so the desired limit can be realized. } as shown in Figure \ref{fig:selfint}.  This places an upper bound $\lambda \leq \lambda_{*}(r_{\textnormal{H}})$ on allowed values of the tension parameter $\lambda$ with sensible Euclidean solutions. 
Explicitly, this upper bound can be found by demanding $2 \Delta z^{\textnormal{ETW}} = \beta$, that is, by enforcing
\begin{equation}
    \beta = 2 \int_{r_{0}^{\textnormal{ETW}}}^{\infty} dr \: \frac{\lambda_{*} r}{L f(r)} \frac{1}{\sqrt{f(r) - \lambda_{*}^{2} r^{2}}} \: .
\end{equation}
A maximal upper bound can be found from $\lambda_{\textnormal{max}} = \max_{r_{\textnormal{H}}} \{ \lambda_{*}(r_{\textnormal{H}}) \}$. 
For example, we find
\begin{itemize}
    \item $d=3$: $\lambda_{\textnormal{max}} L \approx 0.95635$ and $\frac{r_{0}^{\textnormal{ETW}}}{r_{\textnormal{H}}} \lesssim 2.2708$
    \item $d=4$: $\lambda_{\textnormal{max}} L \approx 0.79765$ and $\frac{r_{0}^{\textnormal{ETW}}}{r_{\textnormal{H}}} \lesssim 1.2876$. 
\end{itemize}

\begin{comment}
\begin{figure}
    \centering
    \includegraphics[height=4cm]{SelfIntersection.eps}
    \caption{A well-behaved Euclidean solution and a pathological solution with a self-intersecting brane. Here, $z$ is the angular direction and $r$ is the radial direction; planar directions are suppressed. The ETW brane is shown in red, and the shaded region is included. Self-intersections arise due to the periodicity in $z$ required for the geometry to remain smooth at the coordinate horizon $r = r_{\textnormal{H}}$.}
    \label{fig:selfintersection}
\end{figure}
\end{comment}

\subsection*{Lorentzian picture and cosmology}

In the Lorentzian picture with $z \rightarrow i \zeta$, the ETW brane analytically continues to a spatially flat FRW universe; it is worth noting a few features of the intrinsic geometry of these solutions. 

In terms of the proper time $s$ on the brane defined by
\begin{equation}
    1 = L^{2} f \left( \frac{d \zeta}{ds} \right)^{2} - \frac{1}{f} \left( \frac{dr}{ds} \right)^{2} \: ,
\end{equation}
the metric on the ETW brane is the FRW metric
\begin{equation}
    ds_{d}^{2} = - ds^{2} + r(s)^{2} dx_{\mu} dx^{\mu} \: , \qquad \left( \frac{dr}{ds} \right)^{2} = \lambda^{2} r^{2} - f(r)  \: .
\end{equation}

Comparing to the usual Friedmann equation for a flat universe
\begin{equation}
    \frac{1}{r^{2}} \left( \frac{dr}{ds} \right)^{2} = \frac{8 \pi G \rho}{3}  \: , 
\end{equation}
we infer that our cosmology is effectively sourced by a negative vacuum energy
\begin{equation}
    \frac{8 \pi G \rho_{\Lambda}}{3} = - \frac{1}{L^{2}} \left( 1 - \lambda^{2} L^{2} \right)
\end{equation}
and a ``dark radiation" term
\begin{equation}
    \frac{8 \pi G \rho_{\textnormal{rad}}}{3} = \frac{\mu}{r^{d}} \: .
\end{equation}

We may also note that the total proper time elapsed on the brane is finite, given by
\begin{equation}
    s_{\textnormal{tot}} = 2 \int_{0}^{r_{0}^{\textnormal{ETW}}} \frac{dr}{\sqrt{\lambda^{2} r^{2} - f(r)}} \: .
\end{equation}
That is, the spacetime is geodesically incomplete, beginning with a ``big bang" and ending with a ``big crunch". We thus have that the model introduced here necessarily describes a recollapsing FRW universe with radiation and a negative cosmological constant. 

It was suggested in \cite{Cooper:2018cmb} that locally localized gravity on the ETW brane may be expected in a region which exhibits ``quasistatic" cosmological evolution, and for which the brane remains far outside of the bulk black hole horizon
\begin{equation} \label{eq:cosmologyconditions}
    | H | \ll \frac{1}{L} \: , \qquad r \gg r_{\textnormal{H}} \: ,
\end{equation}
where $H$ is the Hubble parameter. 
Note that we have for the Lorentzian solution
\begin{equation}
    |H| L = \sqrt{ - (1 - \lambda^{2} L^{2}) + \frac{r_{\textnormal{H}}^{d}}{r^{d}}} \: , \qquad \frac{r}{r_{\textnormal{H}}} < \left( 1 - \lambda^{2} L^{2} \right)^{- 1/d} \: ,
\end{equation}
so both conditions require $\lambda L \rightarrow 1$, and therefore lead to self-intersecting solutions in Euclidean signature.

\subsection{Solutions with an interface brane} \label{sec:int}

Analogous to the boundary case in the previous subsection, one may consider a class of models for the gravitational dual of holographic interface conformal field theory (ICFT), 
determined by the Euclidean gravitational action
\begin{equation} \label{eq:action_int}
    \begin{split}
        S & = S_{\textnormal{bulk}} + S_{\textnormal{interface}}^{\textnormal{matter}} \\
        S_{\textnormal{bulk}} & = \frac{1}{16 \pi G_{\textnormal{bulk}}} \sum_{i=1}^{2} \int_{\mathcal{M}_{i}} d^{d+1} x \sqrt{g} \: \left( R - 2 \Lambda_{i} \right) + \frac{1}{8 \pi G_{\textnormal{bulk}}} \int_{\textnormal{interface}} d^{d} y \sqrt{h} \: \left[ K \right] \: ,
    \end{split}
\end{equation}
where we take the brane matter action to be
\begin{equation}
    S_{\textnormal{interface}}^{\textnormal{matter}} = \frac{(1 - d) \kappa}{8 \pi G_{\textnormal{bulk}}} \int_{\textnormal{interface}} d^{d} y \sqrt{ h} \: .
\end{equation}
Here and in the following, the brackets represent the discontinuity $[X] = X_{1} - X_{2}$ across the interface brane separating regions $\mathcal{M}_{1}$ and $\mathcal{M}_{2}$. We are also permitting two different cosmological constants $\Lambda_{i}$, related to the AdS lengths $L_{i}$ as in equation (\ref{eq:cosmoconstant}). 
Here, $\kappa$ lies within the interval
\begin{equation}
    \kappa \in \left( \kappa_{-}, \kappa_{+} \right) \: , \qquad \kappa_{-} = \left| \frac{1}{L_{1}} - \frac{1}{L_{2}} \right| \: , \qquad \kappa_{+} = \frac{1}{L_{1}} + \frac{1}{L_{2}} \: .
\end{equation}

The bulk equations of motion are simply the Einstein equations with the appropriate cosmological constants, while the interface brane trajectory is determined by the junction conditions (see Appendix \ref{app:branetrajectories})
\begin{equation}
    \left[ h_{ab} \right] = 0 \: , \qquad \left[  K_{ab} \right] = \kappa h_{ab} \: .
\end{equation}

We again assume the Euclidean solutions have a $\mathbb{R}^{d-1}$ symmetry; the bulk solutions therefore involve the gluing together of two pieces of the AdS soliton geometry, described by the metric
\begin{equation} \label{eq:interfaceAdSsoliton}
    ds^{2} = L_{i}^{2} f_{i}(r_{i}) dz_{i}^{2} + \frac{dr_{i}^{2}}{f_{i}(r_{i})} + r_{i}^{2} d x_{\mu} dx^{\mu} \: , \qquad f_{i}(r_{i}) = \frac{r_{i}^{2}}{L_{i}^{2}} - \frac{\mu_{i}}{r_{i}^{d-2}} \: ,
\end{equation}
where $L_{i}$ is the AdS radius related to the central charge of the $i^{\textnormal{th}}$ CFT (which we call CFT$_{i}$). One may choose coordinates so that the $x^{\mu}$ agree across the interface joining these two regions; this is our rationale for neglecting a subscript on these coordinates. We may also choose the radial coordinates so that $r_{1} = r_{2} = r$ on the interface, so we will sometimes drop the subscript of $r_{i}$ for quantities on the interface brane. The trajectory of the interface $z_{i}^{\textnormal{int}}(r)$ in each region is determined by equations (4.1) - (4.4) of \cite{May:2021xhz}, which are analogous to (\ref{eq:ETW_traj}) from the ETW brane case. 
These solutions are analyzed extensively in \cite{Simidzija:2020ukv, May:2021xhz}, and we will try to reiterate only the necessary features. 

It will be useful to introduce the parameters 
\begin{equation}
    u = \frac{L_{2}}{L_{1}} \: , \qquad \mu = \frac{\mu_{2}}{\mu_{1}} \: , \qquad e = \frac{\kappa - \kappa_{-}}{\kappa_{+} - \kappa_{-}} \: .
\end{equation}
The full interface solution is then completely specified by the parameters $(L_{1}, \mu_{1}, u, \mu, e)$. 

\subsubsection*{Periodicity of \texorpdfstring{$z_{i}$}{} coordinates in interface solutions}

In contrast to the boundary case in the previous subsection, the coordinate $z_{i}$ need only be taken periodic, with period $\beta_{i}$ given by equation (\ref{eq:zperiod}), if the region $\mathcal{M}_{i}$ includes the coordinate value $r_{i} = r_{\textnormal{H}}^{(i)} = (\mu_{i} L_{i}^{2})^{1/d}$; if not, then the $z_{i}$ coordinate need not be periodic, and in fact the region can be ``multiply wound" from the perspective of this naive periodicity. 

To clarify what we mean by ``multiply wound", we can first define the quantity $\Delta z_{i}^{\textnormal{int}}$ to be equal to the $z_{i}$-coordinate distance traversed by the interface brane from its minimum radius $r_{i}$ to infinity; in equations, we may define
%to be equal to half of the $z_{i}$-coordinate distance subtended by the interface brane; in equations, we may define
\begin{equation} \label{eq:deltaz_int}
    \Delta z_{i}^{\textnormal{int}} \equiv \int_{r_{0}^{\textnormal{int}}}^{\infty} dr_{i} \: \frac{dz_{i}^{\textnormal{int}}}{dr_{i}} \: ,
\end{equation}
where $r_{0}^{\textnormal{int}}$ is the minimum value of both the $r_{1}$ and $r_{2}$ coordinates on the interface brane, and $\frac{dz_{i}^{\textnormal{int}}}{dr_{i}}$ is given by the equation of motion (4.4) in \cite{May:2021xhz}. 
Explicitly, one finds\footnote{The notation $V_{\textnormal{eff}}$ is based on the analysis of \cite{May:2021xhz}, which reduces the dynamics of the interface brane to that of a particle moving an an effective potential. We keep the notation here for consistency.}
\begin{equation} \label{eq:deltaz1}
\begin{split}
    \Delta z_{1}^{\textnormal{int}} = - \frac{1}{L_{1}} \int_{r_{0}^{\textnormal{int}}}^{\infty} \frac{dr}{f_{1} \sqrt{V_{\textnormal{eff}}}} \left( \frac{1}{2 \kappa r} (f_{1} - f_{2}) + \frac{1}{2} \kappa r \right) \: , \quad V_{\textnormal{eff}} = f_{1} - \left( \frac{f_{2} - f_{1} - \kappa^{2} r^{2}}{2 \kappa r} \right)^{2} \: ,
\end{split}
\end{equation}
and an analogous expression for $\Delta z_{2}^{\textnormal{int}}$. 

\begin{figure}
    \centering
    \includegraphics[height=6cm]{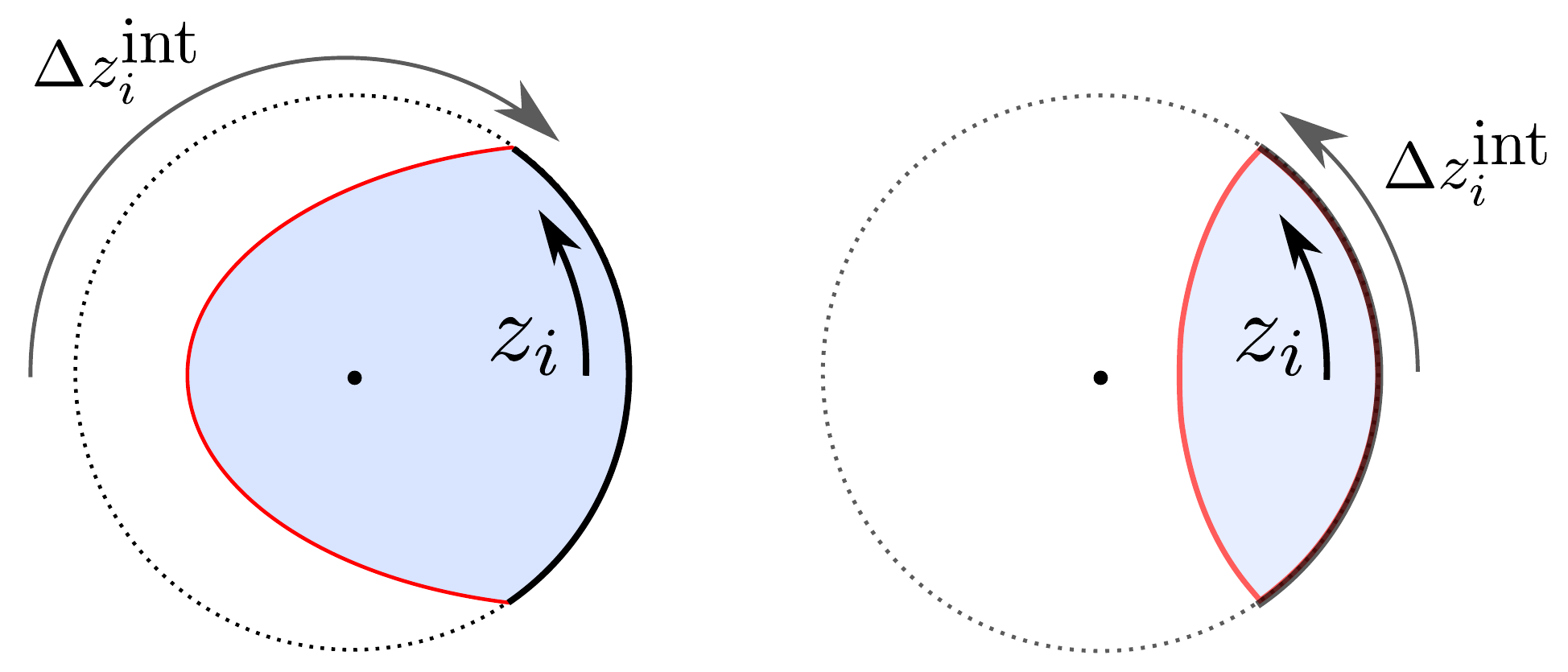}
    \caption{(Left) In the case that $\Delta z_{i}^{\textnormal{int}}<0$, the $i^{\textnormal{th}}$ gravity region includes the horizon. (Right) In the case that $\Delta z_{i}^{\textnormal{int}} > 0$, the $i^{\textnormal{th}}$ gravity region does not include the horizon. }
    \label{fig:deltaz}
\end{figure}

Importantly, $\Delta z_{i}^{\textnormal{int}}$ can be either positive or negative, depending on the data specifying our solution; the former case corresponds to a situation where the $i^{\textnormal{th}}$ gravity region contains the coordinate horizon, whereas the latter case corresponds to a situation where it does not. See Figure \ref{fig:deltaz} for an illustration of this. 

One may then define the quantity $R_{i} = R_{i}(u, \mu, e)$ to be the fraction of the span of the asymptotic $z_{i}$ coordinate in the pure AdS soliton solution (with period $\beta_{i}$) that is covered by the patch associated with CFT$_{i}$ in the interface solution.
We then have two different cases:
\begin{itemize}
    \item If $\Delta z_{i}^{\textnormal{int}}$ is positive, then $R_{i} = \frac{2 \Delta z_{i}^{\textnormal{int}}}{\beta_{i}}$.
    \item If $\Delta z_{i}^{\textnormal{int}}$ is negative, then $R_{i} = 1 - \frac{2 |\Delta z_{i}^{\textnormal{int}}|}{\beta_{i}} = 1 + \frac{2 \Delta z_{i}^{\textnormal{int}}}{\beta_{i}}$.
\end{itemize}
The multiply wound case corresponds to a situation where $\Delta z_{i}^{\textnormal{int}}$ is positive (so that the coordinate horizon is not included), and we have $R_{i} > 1$.

Throughout this work, as a matter of convention, we would like to choose $\mathcal{M}_{1}$ to be the bulk region which excludes $r = r_{\textnormal{H}}$; in this case, $\Delta z_{1}^{\textnormal{int}}$ is positive and $R_{1} = \frac{2 \Delta z_{1}^{\textnormal{int}}}{\beta_{1}}$, while the similarly defined $\Delta z_{2}^{\textnormal{int}}$ is negative and $R_{2} = 1 + \frac{2 \Delta z_{2}^{\textnormal{int}}}{\beta_{2}}$. The condition for this to be the case can be readily derived from checking the sign of the expression (\ref{eq:deltaz1}) for $\Delta z_{1}^{\textnormal{int}}$; one finds that the condition is
\begin{equation} \label{eq:cond_nohorizon}
    \mu < \frac{1}{u^{2}} - \kappa^{2} L_{1}^{2} \: ,
\end{equation}
which we will assume henceforth.

%The authors of \cite{May:2021xhz} observed that a particularly interesting regime in the parameter space occurred for
%\begin{equation}
%    \boxed{u < 1 \: , \qquad \mu < \frac{1}{u} \: , \qquad e \rightarrow 0 } \: ,
%\end{equation}
%where $E_{1}$ was found to exhibit negative energy enhancement. 
%To this end, suppose we restrict to the case that $u < 1$. 

\subsection*{Negative energy enhancement: motivation}

The above Euclidean interface solutions, analytically continued to Lorentzian signature in one of the transverse planar directions, are anticipated to provide a simple holographic description of two CFTs on $\mathbb{R}^{d-2, 1}$ times an interval of width $w_{i}$, coupled at their endpoints via a conformal interface (see the right panel of Figure \ref{fig:sols}). Due to the symmetries of this theory, the energy-momentum tensor must take the form
\begin{equation} \label{eq:cft_T}
    T^{(i)}_{\mu \nu} = \eta_{\mu \nu} \frac{F_{i}}{w_{i}^{d}} \: , \qquad T_{zz}^{(i)} = - \frac{(d-1) F_{i}}{w_{i}^{d}} \: , \qquad T^{(i)}_{\mu z} = 0 \: ,
\end{equation}
where $z$ is the CFT interval direction. Here, $F_{i}$ is a characteristic scale for the vacuum state energy in CFT$_{i}$. 
Following \cite{May:2021xhz}, one may then define
\begin{equation}
    E_{i} = \left( F_{i} / F_{\beta} \right)^{1/d}
\end{equation}
to be the ratio of the scale of the energy density for CFT$_{i}$ on the strip of width $w_{i}$ (in the interface case) to that of the same CFT on a periodic direction of length $\beta = w_{i}$. One expects that this quantity should be a function of the dimensionless ratio 
\begin{equation} \label{eq:def_x}
    x = \frac{w_{2}}{w_{1}} 
\end{equation}
of the widths for the two CFTs. It is useful to note that this ratio is given in terms of bulk quantities by
\begin{equation}
    x = \frac{R_{2} \beta_{2}}{R_{1} \beta_{1}} \: .
\end{equation}

The authors of \cite{May:2021xhz} observed that a particularly interesting regime in the parameter space occurred for\footnote{We note that this limit eventually implies the condition (\ref{eq:cond_nohorizon}), so we need not worry about the latter being satisfied when we are interested in the limit. On the other hand, if we are interested in fixed small $e > 0$, we should check that (\ref{eq:cond_nohorizon}) is still satisfied. }
\begin{equation} \label{eq:NEElimit}
    \boxed{x \: \textnormal{fixed} \: , \qquad u < 1 \: , \qquad e \rightarrow 0 } \: ,
\end{equation}
where the requirement that $x$ remains fixed can be understood as a particular way of taking the limit $\mu \rightarrow 0$, as we will see below.
%\footnote{As we demonstrate below, taking $e \rightarrow 0$ with $x$ fixed will require also sending $\mu \rightarrow 0$.} 
In this limit, $E_{1}$ increases without bound, suggesting that CFT$_{1}$ can exhibit an arbitrarily large negative Casimir energy provided that a family of interfaces realizing this limit can be considered. 
%the significance of this regime will be explained in the following section. 
Interestingly, this effect is only observed in CFT$_{1}$ when $u<1$, i.e. when the central charge of CFT$_{2}$ is smaller than that of CFT$_{1}$.
We will henceforth refer to the limit in (\ref{eq:NEElimit}) as the ``negative energy enhancement" or NEE limit.

In the bulk, this effect can be attributed to the fact that $\mu \rightarrow 0$ corresponds to the limit in which the black hole mass associated to region 1 becomes much larger than that associated to region 2; this results directly in a similar hierarchy for the energy density in the two CFT regions.
We can then think of the NEE limit as taking the lengths $L_{1}, L_{2}$ to be held fixed (as is natural since these correspond to the central charges of the two CFTs), taking the black hole mass $\mu_{1}$ associated with region 1 to be much larger than $\mu_{2}$, and adjusting the interface brane tension as $e \rightarrow 0$ to maintain a fixed value of $x$ in the limit. This relies crucially on the possibility of having a multiply wound region 1, since maintaining fixed $x$ while $\beta_{1} \rightarrow 0$ requires $R_{1} \rightarrow \infty$. 

%and in particular permits $E_{1}$ to become large. For this to be consistent when the lengths $L_{1}, L_{2}$ are held fixed (as is natural since these correspond to the central charges of the two CFTs), we must adjust the brane tension accordingly; 

We may also observe that the limit $e \rightarrow 0$ amounts to shifting the brane out toward the asymptotic region associated to CFT$_{1}$. This can be seen by noting that the Poincar{\'e} angle between the normal to the AdS boundary and the brane in each region is given by \cite{May:2021xhz}
\begin{equation}
    \begin{split}
        \theta_{1} & = \arcsin \left[ \frac{1}{2} \left( \kappa L_{1} + \frac{1}{\kappa L_{1}} - \frac{L_{1}}{\kappa L_{2}^{2}} \right) \right] \stackrel{e \rightarrow 0}{\rightarrow} - \frac{\pi}{2} \: , \\
        \theta_{2} & = \arcsin \left[ \frac{1}{2} \left( \kappa L_{2} + \frac{1}{\kappa L_{2}} - \frac{L_{2}}{\kappa L_{1}^{2}} \right) \right] \stackrel{e \rightarrow 0}{\rightarrow} \frac{\pi}{2} \: .
    \end{split}
\end{equation}

We will now provide some important technical details underlying the above result.

\subsection*{Negative energy enhancement: details}

Since the NEE limit involves fixing $u<1$, we will collect here some important expressions pertaining to this regime. 
Defining
\begin{equation} \nonumber
    \begin{split}
        \alpha_{0} & = \frac{1}{2} \frac{u^{2} (1 - \mu)^{2}}{(1 - u + 2 e u) \sqrt{(1 - \mu u)^{2} + 4 u e \mu (1 - u + u e)} + 2 u (1 + \mu) (1 - e) (1 + e u) - (1 + u) (1 + \mu u)} \\
        \alpha_{1} & = \frac{1}{2} \frac{u(\mu - 1)}{(1 - u)(1 - 2e) - 2 e^{2} u} \\
        \alpha_{2} & = - \frac{(1 - u)(1 - 2 e) - 2 e^{2} u}{\sqrt{e (1 - e)(1 + e u)(1 - u + e u)}} \: ,
    \end{split}
\end{equation}
it was found in \cite{May:2021xhz} that
\begin{equation} \label{eq:R1}
    R_{1} = - \frac{1}{4 \pi} \frac{\alpha_{2}}{\alpha_{0}^{1/d}} \mathcal{I}_{d} \left( \frac{\alpha_{1}}{\alpha_{0}}  , \frac{1}{\alpha_{0}} , \frac{\alpha_{1}^{2} \alpha_{2}^{2}}{4 \alpha_{0}^{2}} \right) + \Theta \left[ \mu - (1 - 2 e) \left( \frac{2}{u} - 1 + 2 e \right) \right] \qquad (\textnormal{for } u < 1)
\end{equation}
where $\Theta( \cdot )$ is a step function and 
\begin{equation}
    \mathcal{I}_{d}(a, b, c) = \int_{1}^{\infty} \frac{dy}{y^{1/d}} \frac{(y-a)}{(y-b) \sqrt{(y-1)(y+c)}} \: .
\end{equation}
Meanwhile, defining
\begin{equation}
    \begin{split}
        \hat{\alpha}_{0} & = \frac{1}{2 u^{2}} \frac{(1 - \frac{1}{\mu})^{2}}{\left( \frac{1}{u} - 1 + 2 e \right) \sqrt{ \left( 1 - \frac{1}{\mu} \frac{1}{u} \right)^{2} + \frac{4e}{\mu} \left( \frac{1}{u} - 1 + e \right) } + 2 \left( 1 + \frac{1}{\mu} \right) \left( 1 - e \right) \left( \frac{1}{u} + e \right) - \left( 1 + \frac{1}{u} \right) \left( 1 + \frac{1}{\mu} \frac{1}{u} \right) } \\
        \hat{\alpha}_{1} & = \frac{1}{2 u^{2}} \frac{\left( 1 - \frac{1}{\mu} \right)}{\left( \frac{1}{u} - 1 \right) \left( \frac{1}{u} + 2 e \right) + 2 e^{2} } \\
        \hat{\alpha}_{2} & = \frac{\left( \frac{1}{u} - 1 \right) \left( \frac{1}{u} + 2 e \right) + 2 e^{2}}{\sqrt{e (1 - e) \left( \frac{1}{u} + e \right) \left( \frac{1}{u} - 1 + e \right)}} \: ,
    \end{split}
\end{equation}
one has
\begin{equation} \label{eq:R2}
    \begin{split}
        R_{2} & = - \frac{1}{4 \pi} \frac{\hat{\alpha}_{2}}{\hat{\alpha}_{0}^{1/d}} \mathcal{I}_{d} \left( \frac{\hat{\alpha}_{1}}{\hat{\alpha}_{0}}, \frac{1}{\hat{\alpha}_{0}}, \frac{\hat{\alpha}_{1}^{2} \hat{\alpha}_{2}^{2}}{4 \hat{\alpha}_{0}^{2}} \right) + \Theta \left[ \frac{1}{\mu} - \left( 1 + 2 e u \right) \left( 2 u - 1 - 2 e u \right) \right] \qquad (\textnormal{for } u < 1) \: .
    \end{split}
\end{equation}
Moreover, the minimum radius of the interface brane is
\begin{equation} \label{eq:r0int} 
\begin{split}
    \left( r_{0}^{\textnormal{int}} \right)^{d} & = \mu_{2} L_{2}^{2} \hat{\alpha}_{0} \qquad (\textnormal{for } u<1) \: .
\end{split}
\end{equation}

Assuming $u < 1$ and $\mu < \frac{1}{u}$ (both of which are prerequisites for the NEE limit), it was found that
\begin{center}
\begin{tabular}{ c c c c c}
 $\alpha_{0} \rightarrow \frac{1 - \mu u}{4e}$ , & $\alpha_{1} \rightarrow \frac{1}{2} \frac{u(\mu - 1)}{1 - u}$ ,  & $\alpha_{2} \rightarrow - \sqrt{\frac{1-u}{e}}$ , & & $(u < 1 \: , \: \mu < \frac{1}{u} \: , \: e \rightarrow 0)$ \\ 
 $\hat{\alpha}_{0} \rightarrow \frac{1 - \mu u }{4 e \mu u^{2}} $ , & $\hat{\alpha}_{1} \rightarrow \frac{1}{2\mu} \frac{\mu - 1}{ 1 - u } $ ,  & $\hat{\alpha}_{2} \rightarrow \frac{1}{u} \sqrt{\frac{ 1 - u }{e}}$ . & & $(u < 1 \: , \: \mu < \frac{1}{u} \: , \: e \rightarrow 0)$
\end{tabular}
\end{center}
Thus, defining
\begin{equation}
    \mathcal{I}_{0} = \frac{4^{1/d}}{4 \pi} \int_{1}^{\infty} \frac{dy}{y^{1/d} \sqrt{y(y-1)}} = \frac{4^{1/d}}{4 \pi} \frac{ \Gamma \left( \frac{1}{2} \right) \Gamma \left( \frac{1}{d} \right) }{\Gamma \left( \frac{1}{2} + \frac{1}{d} \right)} \: ,
\end{equation}
it was found that
\begin{equation} \label{eq:asymptotic_R1R2}
    \begin{split}R_{1} & = \frac{2 \Delta z_{1}}{\beta_{1}} \sim  \frac{\mathcal{I}_{0}}{e^{\frac{1}{2} - \frac{1}{d}}} \frac{\sqrt{1 - u}}{(1 - \mu u)^{1/d}} \qquad \qquad \qquad \qquad \qquad \qquad (u < 1 \: , \: \mu < \frac{1}{u} \: , \: e \rightarrow 0)  \\
    R_{2} & = 1 + \frac{2 \Delta z_{2}}{\beta_{2}} \sim 1 - \mu^{1/d} u^{2/d - 1} \frac{\mathcal{I}_{0}}{e^{1/2 - 1/d}} \frac{\sqrt{1-u}}{\left( 1 - \mu u \right)^{1/d}} \: . \qquad \:  (u < 1 \: , \: \mu < \frac{1}{u} \: , \: e \rightarrow 0)
    \end{split}
\end{equation}
Moreover, in this limit, the minimum radius goes as
\begin{equation} \label{eq:asymptotic_r0int}
    r_{0}^{\textnormal{int}}  \sim r_{\textnormal{H}}^{(2)} \left( \frac{1 - u \mu}{4 e \mu u^{2}} \right)^{1/d} =  r_{\textnormal{H}}^{(1)} \left( \frac{1 - u \mu}{4 e } \right)^{1/d} \: . \qquad \qquad \qquad (u < 1 \: , \: \mu < \frac{1}{u} \: , \: e \rightarrow 0)
\end{equation}

So far, we have been considering limits with a general fixed value of $\mu$; eventually, we would like to instead consider the NEE limit in which we instead fix $x$. 
Indeed, it is clear from the expression for $R_{2}$ that we cannot consistently take $u<1$ and $\mu < \frac{1}{u}$ fixed and send $e \rightarrow 0$; doing so would result in a negative value of $R_{2}$, taking the result beyond its regime of validity. It is therefore more convenient to express the results in terms of the ratio $x$ defined in (\ref{eq:def_x}), which is related to $\mu(x)$ at leading order by
\begin{equation}
    \mu(x) = \frac{e^{\frac{d}{2} - 1}}{u^{2}} \left( \frac{u}{(1+x) \mathcal{I}_{0} \sqrt{1 - u}} \right)^{d} \: .
\end{equation}
The NEE limit properly involves fixing $u< 1$ and $x$, and sending $e \rightarrow 0$, which will also send $\mu \rightarrow 0$ as a result of this equation. 
The authors of \cite{May:2021xhz} then found
\begin{equation}
    E_{1} \sim \frac{1}{e^{\frac{1}{2} - \frac{1}{d}}} \mathcal{I}_{0} \sqrt{1 - u} \quad \textnormal{and} \quad E_{2} \sim \frac{x}{1+x} \: . \qquad \qquad \qquad \qquad \text{(NEE)}
\end{equation}
This limit
%result 
is the most physical from the CFT perspective, since one would typically like to keep the dimensions of the strip on which the CFTs are defined fixed while varying a parameter related to properties of the conformal interface.

\section{Bottom-up model with constant tension branes} \label{sec:constantt}

\begin{figure}
    \centering
    \includegraphics[height=5cm]{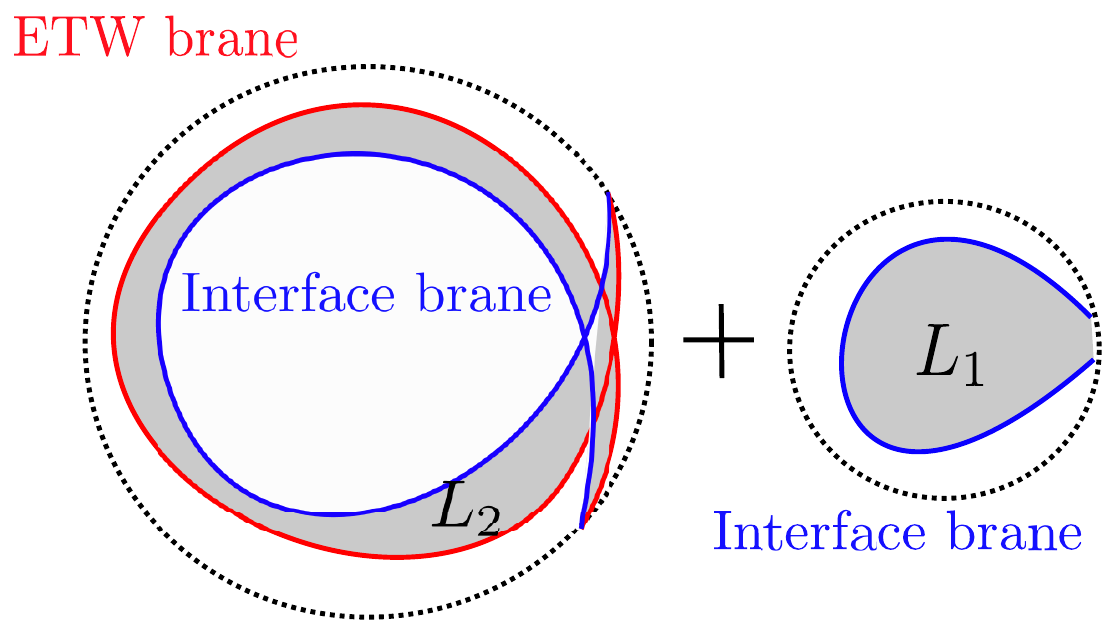}
    \caption{Two Euclidean AdS soliton regions of a holographic interface solution. Here, $z_{i}$ is the angular direction and $r_{i}$ is the radial direction, with $i=1$ on the left and $i=2$ on the right; planar directions are suppressed. In this figure, region 1 is ``multiply wound" in the $z_{1}$ direction, while region 2 (which includes the horizon $r_{2} = r_{\textnormal{H}}$) is not.  }
    \label{fig:multiplywound}
\end{figure}

In the previous section, we reviewed a model of holographic BCFT and its application to cosmology, as well as a model for holographic interfaces exhibiting an interesting ``negative energy enhancement" effect in an appropriate limit. In this section, we would like to combine these two models, considering a gravitational bulk with both an ETW brane and an interface brane; see Figure \ref{fig:multiplywound}. The motivation for this is to see whether, in this augmented model, it is possible to obtain well-behaved Euclidean solutions, without intersecting or self-intersecting branes, so that conditions analogous to those of (\ref{eq:cosmologyconditions}) hold for the ETW brane cosmology arising in the Lorentzian continuation. 

%Presented holographic duals with ETW and interface branes. Would now like to combine these ingredients, look for a Euclidean solution with both types of brane. ssss

%As reviewed in the previous section, in a simple model of holographic CFT interfaces with a constant tension interface brane, one can obtain enhanced negative energies in one of the CFTs in an appropriate ``critical tension" limit. It is interesting to consider how this result may be adapted to a situation with an additional ETW brane in one of the bulk regions, so that we now have a bulk dual of BCFT with two thin branes (an ETW brane and an interface brane); see Figure \textbf{Fig}. 
We might expect that an effective description of the physics in this combined model should involve a non-gravitational CFT joined at an interface to a CFT coupled to gravity;\footnote{This is the usual Karch/Randall/Sundrum mechanism \cite{Randall:1999vf}: given a holographic CFT, we anticipate that introducing a ``UV" or ``Planck" brane in the bulk has the effect of introducing a cutoff to the CFT and coupling to dynamical gravity. Here, we anticipate that introducing an ETW brane in region 1 has this effect on CFT$_{1}$, while region 2 is not cut off and CFT$_{2}$ therefore does not couple to gravity directly. See \cite{VanRaamsdonk:2021qgv} for a discussion of this.} the background for the latter is the geometry of the ETW brane in the bulk picture, which can be interpreted as a traversable wormhole.\footnote{We are here thinking about the Lorentzian picture where we Wick rotate one of the $x^{\mu}$ coordinates.} 
It has been argued that large quantities of negative null energy would be required to support such traversable wormholes \cite{Freivogel:2019lej}; as pointed out in \cite{VanRaamsdonk:2021qgv}, it is therefore natural to look for bulk solutions with both an ETW brane and an interface brane in the critical interface tension or NEE limit considered in the previous section. We will argue in this section that it is not possible to find such solutions in that limit in the present model, prompting a modification to the model explored in the following section. 
%After making some brief comments, we leave a complete exploration of the parameter space in this model to future work.

Before preceding to elucidate this result, it is worth taking a moment to comment on the anticipated effect of adding an interface brane to our model, beyond what we have already mentioned. Introducing this additional ingredient into our model allows us to describe a larger class of holographic duals of boundary states, with different boundary spectra, each of which will give rise to different effective theories on the brane. 
The geometry of the region between the interface and ETW branes is intimately related to the physics of the BCFT boundary degrees of freedom, including the number of these degrees of freedom; this can be understood as an example of generalized wedge holography \cite{Akal:2020wfl}, where we have a bulk dual of a BCFT involving two ``wedges" of AdS separated by an interface brane (see also \cite{VanRaamsdonk:2021duo} for a microscopic version of this phenomenon).
%\footnote{No precise distinction between ambient and boundary degrees of freedom in BCFT, and gravitating regions they correspond to.} 
In particular, as in the previous section, we are typically interested in the case $u<1$, so that, despite considering a BCFT with central charge $c_{2}$, we have a holographic dual including a spacetime region which we expect to be described by a CFT with larger central charge $c_{1} > c_{2}$; this suggests that the corresponding BCFT is defined by permitting many degrees of freedom localized near the boundary, and we anticipate that, as a result, the effective theory on the brane will also have more degrees of freedom than the non-gravitating CFT in the effective picture.

One could nominally be concerned that adding an interface brane could disrupt the condition for gravity localization, namely %that having $r_{0}^{\textnormal{ETW}} \gg r_{\textnormal{H}}^{(1)}$, or 
an ETW brane far in the UV; for example, one could worry that the interface brane may localize gravity in this setup.
%should be sufficient. 
However, we do not expect that the interface brane should interfere with the gravity localization condition, particularly in a region where the ETW brane is much further in the UV than the interface brane. While it is true that interface branes can also exhibit gravity localization (as in the original Randall-Sundrum II model \cite{Randall:1999vf}), we have in our case a situation where the interface brane is not situated at a local maximum of the warp factor, and we therefore do not expect it to support a localized bound state of the $(d+1)$-dimensional graviton. Moreover, following the intuition of \cite{Karch:2000ct}, we can observe that the ETW brane localization phenomenon should be a consequence of local physics, rather than depending on global features of the bulk spacetime; provided we are interested in a region where the ETW brane and interface brane are significantly separated, we should be able to recover locally localized gravity. Just as in \cite{Karch:2000ct}, we expect to find a massive, normalizable Kaluza-Klein mode whose wavefunction localizes to the ETW brane, but whose precise profile depends on the details of the IR physics, including the location and geometry of the interface brane.

\subsection{Non-existence of solutions}

%We would now like to consider a model involving both an ETW brane and an interface brane; t
The Euclidean action for the theory considered in this section is obtained by straightforwardly combining those for the two models considered in the previous section, found in (\ref{eq:action_bound}) and (\ref{eq:action_int}), and is given in Appendix \ref{app:branetrajectories}. 
We assume without loss of generality that the ETW brane is added to region 1, so that in the effective description, CFT$_{1}$ is coupled to gravity via the Randall/Sundrum mechanism while CFT$_{2}$ is not. 

We again consider Euclidean solutions with $\mathbb{R}^{d-1}$ symmetry (or $\mathbb{R}^{d-2, 1}$ symmetry upon Wick rotating one of the $x^{\mu}$ coordinates); these are again pieces of the Euclidean AdS soliton geometry, which we will continue to parametrize as in (\ref{eq:interfaceAdSsoliton}). 
The interface brane trajectory in the two regions is given by the same equation of motion for $z_{i}^{\textnormal{int}}(r)$ as in Section \ref{sec:int}, and $\Delta z_{i}^{\textnormal{int}}$ still denotes the $z_{i}$-coordinate distance traversed by the interface brane from its minimum radius $r_{i} = r_{0}^{\textnormal{int}}$ to infinity, as in (\ref{eq:deltaz_int}); $z_{1}^{\textnormal{ETW}}(r_{1})$ and $\Delta z_{1}^{\textnormal{ETW}}$ are analogous quantities for the ETW brane, following the definitions in (\ref{eq:ETW_traj}) and (\ref{eq:deltaz_ETW}). We will assume $z_{1}^{\textnormal{int}}(r_{0}^{\textnormal{int}}) = 0$ without loss of generality, a choice for the zero of the coordinate $z_{1}$; solutions where the ETW brane and interface brane join properly at infinity must therefore have $z_{1}^{\textnormal{ETW}}(r_{0}^{\textnormal{ETW}}) = 0$ by symmetry, so we will assume this in the following.

As in the previous section, we will be interested in the case that region 1 does not include the coordinate horizon $r_{1} = r_{\textnormal{H}}^{(1)}$ while region 2 does include the coordinate horizon $r_{2} = r_{\textnormal{H}}^{(2)}$; this permits region 1 to be multiply wound, which is what we expect to be required to obtain the negative energy enhancement effect in CFT$_{1}$. Recall that this implies $\Delta z_{1} > 0$, and therefore $R_{1} = \frac{2 \Delta z_{1}}{\beta_{1}} > 0$. 

\subsection*{Conditions for existence of solutions}

It is clear that solutions of the desired type, parametrized by $(L_{1}, \mu_{1}, u, \mu, e)$ and the ETW brane tension $\lambda$ (and with $z_{1}^{\textnormal{int}}(r_{0}^{\textnormal{int}}) = z_{1}^{\textnormal{ETW}}(r_{0}^{\textnormal{ETW}}) = 0$ as mentioned above), will exist if and only if the following conditions are satisfied:
\begin{enumerate}
    \item $R_{2}(u, \mu, e) > 0 \: ;$
    \item $\Delta z_{1}^{\textnormal{int}} = \Delta z_{1}^{\textnormal{ETW}} \: ;$
    \item $r_{0}^{\textnormal{ETW}} > r_{0}^{\textnormal{int}}$ and $|z_{1}^{\textnormal{int}}(r_{1})| > |z_{1}^{\textnormal{ETW}}(r_{1})|$ for all $r_{1} > r_{0}^{\textnormal{ETW}}$. 
\end{enumerate}
The first condition ensures that the interface solution on its own would be well-defined\footnote{The requirement $R_{1} > 0$ is already enforced by our assumption $\Delta z_{1} > 0$.} (the width of CFT$_{2}$ is non-negative), the second that the ETW brane and interface brane join properly (they subtend the same $z_{1}$-coordinate length), and the third that the ETW brane always sits at a larger value of the radial coordinate than the interface brane in region 1. 

In particular, to demonstrate the \textit{non-existence} of solutions for a given set of parameters $(L_{1}, \mu_{1}, u, \mu, e)$ and \textit{any} $\lambda$, it is sufficient to show that one of the following two conditions is not satisfied:
\begin{itemize}
    \item[\textbf{(C1)}] $R_{2}(u, \mu, e) > 0$
    \item[\textbf{(C2)}] For $\lambda = \lambda_{0}$ with $\lambda_{0}$ defined by $f_{1}(r_{0}^{\textnormal{int}}) = \lambda_{0}^{2} (r_{0}^{\textnormal{int}})^{2}$, one has
    \begin{equation}
        \frac{\Delta z_{1}^{\textnormal{ETW}}}{\Delta z_{1}^{\textnormal{int}}} < 1 \: .
    \end{equation}
\end{itemize}

The latter condition requires a brief explanation. Here, $\lambda_{0}$ is the value of the ETW brane tension $\lambda$ for which the minimum radius $r_{0}^{\textnormal{ETW}}$ of the ETW brane would coincide with that of the interface brane, $r_{0}^{\textnormal{int}}$. We know from (\ref{eq:r0ETW}) that $r_{0}^{\textnormal{ETW}}$ monotonically increases over $(r_{\textnormal{H}}^{(1)}, \infty)$ as a function of $\lambda L_{1} \in (0, 1)$, and as shown in Appendix \ref{app:monotonicity}, we have $\Delta z_{1}^{\textnormal{ETW}}$ monotonically increasing from zero to infinity over the same range of ETW brane tensions. Consequently, condition (C2) above is equivalent to the existence of a tension $\lambda L_{1} \in (\lambda_{0} L_{1}, 1)$ such that
\begin{equation}
    r_{0}^{\textnormal{ETW}} > r_{0}^{\textnormal{int}} \: , \qquad \Delta z_{1}^{\textnormal{ETW}}  = \Delta z_{1}^{\textnormal{int}} \: .
\end{equation}

In the following, we will 
%first 
show that these two conditions cannot simultaneously be satisfied in the NEE limit.
%critical tension limit $e \rightarrow 0$. Subsequently, we will extend this result to show that they are inconsistent for any choice of parameters $(u, \mu, e)$ satisfying $u < 1$. 

%We first show that it is not possible to realize this limit

%We then extend this observation to the entire region of parameter space $u < 1$

%Give two conditions, and reduce to expressions involving $u, \mu, e$

\subsection*{No solutions in the NEE limit}

We can begin by determining when (C2) can be satisfied. Recalling the limiting behaviour of (\ref{eq:asymptotic_R1R2}) and (\ref{eq:asymptotic_r0int})
\begin{equation}
    \Delta z_{1}^{\textnormal{int}} \sim \frac{2 \pi L_{1}}{ d r_{\textnormal{H}}^{(1)}} \frac{\mathcal{I}_{0}}{e^{\frac{1}{2} - \frac{1}{d}}} \frac{\sqrt{1-u}}{(1 - \mu u)^{1/d}} \: , \qquad r_{0}^{\textnormal{int}} \sim \frac{r_{\textnormal{H}}^{(1)} (1 - u \mu)^{1/d}}{(4 e)^{1/d}} \: ,
\end{equation}
we have from the definition of $\lambda_{0}$
\begin{equation}
    \lambda_{0} = \frac{1}{L_{1}} \left( 1 - \frac{2e}{(1 - \mu u)} \right) + O(e^{2}) \: .
\end{equation}
In the limit $e \rightarrow 0$,
\begin{equation}
\begin{split}
    \Delta z_{1}^{\textnormal{ETW}}(\lambda = \lambda_{0}) % & \sim \frac{L }{\sqrt{\frac{4e}{(1 - \mu u)}}} \int_{r_{0}}^{\infty} dr  \: \frac{r^{\frac{3d}{2} - 2}}{ (r^{d} - r_{\textnormal{H}}^{d} )} \frac{1}{\sqrt{ r^{d}   - r_{0}^{d}}} \\
    & \sim \frac{2 \pi L_{1}}{d r_{\textnormal{H}}^{(1)}} \left( \frac{1 - \mu u}{e} \right)^{\frac{1}{2} - \frac{1}{d}} \mathcal{I}_{0} \: ,
\end{split}
\end{equation}
and thus
\begin{equation}
    \begin{split}
        \frac{\Delta z_{1}^{\textnormal{ETW}}(\lambda = \lambda_{0})}{\Delta z_{1}^{\textnormal{int}}} \sim \sqrt{\frac{1 - \mu u}{1 - u}} \: .
    \end{split}
\end{equation}

Assuming fixed $u<1$, we thus have two possibilities.
If $\mu < 1$, then this quantity will be greater than one, so that (C2) is not satisfied in the limit, while if $1 < \mu < \frac{1}{u}$, then this quantity will be less than one, so (C2) is satisfied and a solution may exist.

On the other hand, we have already seen in (\ref{eq:asymptotic_R1R2}) that
\begin{equation}
    R_{2} \sim 1 - \mu^{1/d} u^{2/d - 1} \frac{\mathcal{I}_{0}}{e^{\frac{1}{2} - \frac{1}{d}}} \frac{\sqrt{1 - u}}{(1 - \mu u)^{1/d}} \: ;
\end{equation}
for fixed $u<1$, we see that requiring $R_{2} > 0$ in the $e \rightarrow 0$ limit requires $\mu \rightarrow 0$. Thus, the condition $\mu \rightarrow 0$ imposed by (C1) is inconsistent with $\mu > 1$ imposed by (C2).

\subsection*{Solutions for \texorpdfstring{$u < 1$}{}}

While we have shown that it is not possible to obtain solutions with an ETW brane and an interface brane that join properly in the NEE limit, it is certainly the case that well-behaved solutions exist elsewhere in the parameter space. 
The reason that we are not concerned with these solutions here is that they are not expected to be relevant for cosmology, on the basis of arguments we have previously mentioned regarding the effective description of the bulk physics of this model; without the NEE limit, we expect the background for the gravitational CFT to have a 4D curvature scale $L_{4}$ of order $L_{\textnormal{Planck}}$ (the cutoff scale for the gravitational CFT) rather than some hierarchically larger length scale.\footnote{As observed in equation (4.10) 
of \cite{VanRaamsdonk:2021qgv}, the boundary central charge $c_{\textnormal{3D}} = L_{4}^2/G_{4}$ in our setup, which is the bulk description of a holographic BCFT, is equal (up to $O(1)$ factors) to the coefficient $F$ of the energy density for the gravitational CFT in an expression analogous to (\ref{eq:cft_T}), i.e. in $T_{00} \sim F/w^{4}$. One would expect that the typical 
value for $F$ is roughly equal to the number of degrees of freedom in the gravitational CFT, which is not expected to be large in general,
%for a realistic model, 
implying that we should generically expect $L_{4} \sim L_{\textnormal{Planck}}$
%. So to get a 4D spacetime where $L_4^2/G_4$ is large, we want an enhancement of the energy by a factor of $c_3D/c_0$ relative to the  typical scale. This can occur in the critical tension or 
unless we consider something like the NEE limit. %, but would not be expected otherwise. 
We thank Mark Van Raamsdonk for emphasizing this point.} 
%allow $\frac{r_{0}^{\textnormal{ETW}}}{ r_{\textnormal{H}}^{(1)}} \gg 1$, as required for gravity localization in the cosmology. 
Nonetheless, we briefly comment here about the larger parameter space.
%it may be interesting to determine the largest hierarchy $\frac{r_{0}^{\textnormal{ETW}}}{ r_{\textnormal{H}}^{(1)}}$ that can be obtained in this model, and where in the parameter space this occurs. 
%since such a hierarchy is required to ensure that something like the conditions (\ref{eq:cosmologyconditions}) for gravity localization in Lorentzian signature hold. 

A convenient feature for an investigation of this parameter space is that both conditions (C1) and (C2) can be expressed in terms of inequalities which depend only on the parameters $(u, \mu, e)$. 
From (\ref{eq:R2}), we recall that, if $u<1$, the first condition yields the inequality
\begin{equation}
    \begin{split}
        R_{2}(u, \mu, e) & = R_{1} \left( \frac{1}{u} , \frac{1}{\mu}, e \right)  = 1 - \frac{1}{4 \pi} \frac{\hat{\alpha}_{2}}{\hat{\alpha}_{0}^{1/d}} \mathcal{I}_{d} \left( \frac{\hat{\alpha}_{1}}{\hat{\alpha}_{0}}, \frac{1}{\hat{\alpha}_{0}}, \frac{\hat{\alpha}_{1}^{2} \hat{\alpha}_{2}^{2}}{4 \hat{\alpha}_{0}^{2}} \right) > 0 \qquad (u < 1) \: .
    \end{split}
\end{equation}

On the other hand, recalling from (\ref{eq:r0int}) that
\begin{equation}
    \frac{(r_{0}^{\textnormal{int}})^{d}}{\mu_{1} L_{1}^{2}} = \mu u^{2} \hat{\alpha}_{0} \qquad (u<1) \: ,
\end{equation}
and from (\ref{eq:r0ETW}) that 
\begin{equation}
    \lambda = \frac{\sqrt{f_{1}(r_{0}^{\textnormal{ETW}})}}{r_{0}^{\textnormal{ETW}}} = \frac{1}{L_{1}} \sqrt{1 - \frac{\mu_{1} L_{1}^{2}}{(r_{0}^{\textnormal{ETW}})^{d}}} \: , \qquad r_{0}^{\textnormal{ETW}} = \left( \frac{\mu_{1} L_{1}^{2}}{1 - L_{1}^{2} \lambda^{2}} \right)^{1/d} \: ,
\end{equation}
we see that when the tension takes the value $\lambda_{0}$ for which $r_{0}^{\textnormal{ETW}} = r_{0}^{\textnormal{int}} = r_{0}$, we have
\begin{equation}
    \begin{split}
        \Delta z_{1}^{\textnormal{ETW}} & = \int_{r_{0}}^{\infty} dr \: \frac{r \lambda_{0}}{L_{1} f_{1}(r)} \frac{1}{\sqrt{f_{1}(r) - r^{2} \lambda_{0}^{2}}} \\
        %& = \frac{L_{1}^{2} \lambda_{0}}{d r_{0} \sqrt{1 - L_{1}^{2} S_{0}^{2}}} \int_{1}^{\infty} y^{-\frac{1}{d}} dy \: \frac{ y }{ \left( y - \frac{\mu_{1} L_{1}^{2} }{r_{0}^{d}} \right)} \frac{1}{\sqrt{ y(y - 1) }} \\
        & = \frac{L_{1} \sqrt{\mu u^{2} \hat{\alpha}_{0} - 1} }{d r_{0} } \mathcal{I}_{d} \left( 0, \frac{1}{\mu u^{2} \hat{\alpha}_{0}} , 0 \right) \: .
    \end{split}
\end{equation}
%where we used 
%\begin{equation}
%    y = \left( r / r_{0} \right)^{d} \: , \qquad \frac{1}{d} \frac{dy}{y} = \frac{dr}{r} \: .
%\end{equation}
We therefore have
\begin{equation}
    \begin{split}
        \frac{\Delta z_{1}^{\textnormal{ETW}}(\lambda = \lambda_{0})}{\Delta z_{1}^{\textnormal{int}}} %& = \frac{2 \Delta z_{\textnormal{ETW}}(S_{0})}{\beta_{1} R_{1}(u, \mu, e)} = \frac{d}{2 \pi} \frac{(\mu_{1} L_{1}^{2})^{1/d}}{L_{1}} \frac{\Delta z_{\textnormal{ETW}}}{R_{1}(u, \mu, e)} \\
        & = - 2 \left( \frac{\alpha_{0}}{\mu u^{2} \hat{\alpha}_{0}} \right)^{1/d}  \sqrt{ \mu u^{2} \hat{\alpha}_{0} - 1} \frac{\mathcal{I}_{d}(0, \frac{1}{\mu u^{2} \hat{\alpha}_{0}}, 0)}{\alpha_{2} \mathcal{I}_{d} \left( \frac{\alpha_{1}}{\alpha_{0}}, \frac{1}{\alpha_{0}}, \frac{\alpha_{1}^{2} \alpha_{2}^{2}}{4 \alpha_{0}^{2}} \right)} \: .
    \end{split}
\end{equation}
%where
%\begin{equation}
%    \begin{split}
%        \alpha_{0} & = \frac{1}{2} \frac{u^{2} (1 - \mu)^{2}}{\left( 1 - u + 2 e u \right) \sqrt{ \left( 1 - \mu u \right)^{2} + 4 e u \mu \left( 1 - u + e u \right) } + 2 u \left( 1 + \mu \right) \left( 1 - e \right) \left( 1 + e u \right) - \left( 1 + u \right) \left( 1 + u \mu \right) } \\
%        \alpha_{1} & = \frac{1}{2 } \frac{u \left( \mu - 1 \right)}{\left( 1 - u \right) \left( 1 - 2e \right) - 2 e^{2} u } \\
%        \alpha_{2} & = - \frac{\left( 1 - u \right) \left( 1 - 2e \right) - 2 e^{2} u}{\sqrt{e (1 - e) \left( 1 + eu \right) \left( 1 - u + e u \right)}} \: .
%    \end{split}
%\end{equation}

It follows that we can express the conditions introduced above as
\begin{itemize}
    \item[\textbf{(C1)}] $1 - \frac{1}{4 \pi} \frac{\hat{\alpha}_{2}}{\hat{\alpha}_{0}^{1/d}} \mathcal{I}_{d} \left( \frac{\hat{\alpha}_{1}}{\hat{\alpha}_{0}}, \frac{1}{\hat{\alpha}_{0}}, \frac{\hat{\alpha}_{1}^{2} \hat{\alpha}_{2}^{2}}{4 \hat{\alpha}_{0}^{2}} \right) > 0$ 
    \item[\textbf{(C2)}] $- 2 \left( \frac{\alpha_{0}}{\mu u^{2} \hat{\alpha}_{0}} \right)^{1/d}  \sqrt{ \mu u^{2} \hat{\alpha}_{0} - 1} \frac{\mathcal{I}_{d}(0, \frac{1}{\mu u^{2} \hat{\alpha}_{0}}, 0)}{\alpha_{2} \mathcal{I}_{d} \left( \frac{\alpha_{1}}{\alpha_{0}}, \frac{1}{\alpha_{0}}, \frac{\alpha_{1}^{2} \alpha_{2}^{2}}{4 \alpha_{0}^{2}} \right)} < 1$. 
\end{itemize}
These expressions are a convenient reformulation of (C1) and (C2) for the purposes of verifying their compatibility within the parameter space.

As a preliminary for determining where such well-behaved solutions could exist in the parameter space, our goal in the remainder of this section will be to indicate a portion of the parameter space where these solutions cannot occur. We will restrict our attention to the region satisfying:
\begin{itemize}
    \item $u < 1$
    \item $\mu < \textnormal{min}\{ \frac{1}{u},  \frac{1}{u^{2}} - \kappa^{2} L_{1}^{2}\}$;
\end{itemize}
however, one could ultimately explore the parameter space more broadly. 
We note that, together, these conditions imply $0 < e < \frac{1}{2}$. We therefore assume here that
\begin{equation}
    \boxed{ 0 < u < 1 \: , \quad 0 < e < \frac{1}{2} \: , \quad \mu < \textnormal{min} \Big\{ \frac{1}{u}, \frac{1}{u^{2}} \left( 1 - (1 - u + 2eu)^{2} \right)  \Big\}  } \: .
\end{equation}

We will denote 
\begin{equation}
\begin{split}
    c_{1}(u, \mu, e) & = \frac{1}{4 \pi} \frac{\hat{\alpha}_{2}}{\hat{\alpha}_{0}^{1/d}} \mathcal{I}_{d} \left( \frac{\hat{\alpha}_{1}}{\hat{\alpha}_{0}}, \frac{1}{\hat{\alpha}_{0}}, \frac{\hat{\alpha}_{1}^{2} \hat{\alpha}_{2}^{2}}{4 \hat{\alpha}_{0}^{2}} \right) \: , \\
    c_{2}(u, \mu, e) & = - 2 \left( \frac{\alpha_{0}}{\mu u^{2} \hat{\alpha}_{0}} \right)^{1/d}  \sqrt{ \mu u^{2} \hat{\alpha}_{0} - 1} \frac{\mathcal{I}_{d}(0, \frac{1}{\mu u^{2} \hat{\alpha}_{0}}, 0)}{\alpha_{2} \mathcal{I}_{d} \left( \frac{\alpha_{1}}{\alpha_{0}}, \frac{1}{\alpha_{0}}, \frac{\alpha_{1}^{2} \alpha_{2}^{2}}{4 \alpha_{0}^{2}} \right)} \: , 
\end{split}
\end{equation}
so that the condition (C$_{i}$) corresponds to the inequality $c_{i}(u, \mu, e) < 1$.

\begin{figure}
    \centering
    \includegraphics[height=8cm]{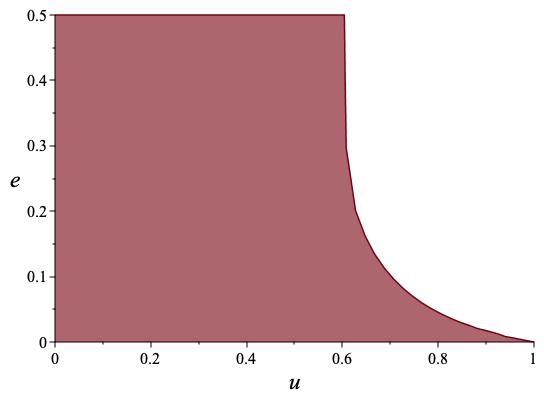}
    \caption{Plot of ``ruled out" region of the $(u, e)$-plane. Here, the region shaded in red is part of the parameter space where we \textit{do not} expect solutions to occur, as conditions (C1) and (C2) cannot be simultaneously satisfied. The remaining unshaded region in the upper right corner may or may not have solutions (our procedure for ruling out regions of the parameter space was not exhaustive). }
    \label{fig:ruledout}
\end{figure}

We observe (but will not attempt to prove here) that, for fixed $(e, u)$, the function $c_{1}(u, \mu, e)$ is monotonically decreasing in $\mu$, while $c_{2}(u, \mu, e)$ is monotonically increasing in $\mu$. Assuming that this is true, then a pair of parameters $(u, e)$ may be ruled out, meaning that they do not permit a well-behaved solution, if the solution $\mu = \mu_{0}$ to the equation $c_{1}(u, \mu, e) = 1$ (which we may obtain numerically) yields $c_{2}(u, \mu_{0}, e) > 1$. Using this approach, we construct the plot shown in Figure \ref{fig:ruledout}. The shaded portion of the plot corresponds to a region of the parameter space which has been ruled out, meaning that it does not contain any well-behaved solutions; the unshaded portion may or may not contain solutions (further investigation would be needed to determine this). This plot already confirms the conclusion of Section \ref{sec:constantt} that solutions cannot exist in the NEE limit, which requires $e \rightarrow 0$ for fixed $u < 1$.

\section{Bottom-up model with Einstein-Hilbert term on the ETW brane} \label{sec:EH}

We will now consider a generalization of the model considered above, where an Einstein-Hilbert term is added to the ETW brane.\footnote{We do not add an Einstein-Hilbert term to the interface brane, as this complicates the analysis, though we provide the relevant equations in Appendix \ref{app:branetrajectories}. } In particular, we modify the 
%interface and 
ETW brane contribution to the action of the previous section to become
\begin{equation}
    \begin{split}
        %S_{\textnormal{interface}} & = \frac{1}{16 \pi G_{\textnormal{interface}}} \int_{\textnormal{interface}} d^{d} y \sqrt{h} \: R^{(d)} + S_{\textnormal{interface}}^{\textnormal{matter}} \\
        S_{\textnormal{ETW}} & = \frac{1}{16 \pi G_{\textnormal{ETW}}} \int_{\textnormal{ETW}} d^{d} y \sqrt{h} \: R^{(d)} + S_{\textnormal{ETW}}^{\textnormal{matter}} \: ,
    \end{split}
\end{equation}
where the matter contributions are from constant tension terms as before, and where we will introduce the constant $\gamma$ defined by
\begin{equation}
    \frac{1}{G_{\textnormal{ETW}}} = \frac{\gamma}{G_{\textnormal{bulk}}} \: .
\end{equation}
Again, for the solutions with the desired symmetry, the bulk consists of two AdS soliton regions; the equations of motion for the ETW brane can be found in Appendix \ref{app:EHeqs}. 

%From the equations of motion, one finds that the ETW brane trajectory is determined by (see Appendix APP)
%\begin{equation}
%    \frac{d z_{1}^{\textnormal{ETW}}}{dr} = \frac{\sqrt{(d-2) \gamma \lambda f_{1}(r) + 2 \lambda^{2} r^{2} - f_{1}(r) + \frac{f_{1}(r)}{r} \sqrt{(d-2)^{2} \gamma^{2} f_{1}(r) + \left( 2 (d-2) \gamma \lambda + 1 \right) r^{2} }}}{\sqrt{2} L_{1} f_{1}(r) \sqrt{ f_{1}(r) - r^{2} \lambda^{2}}} \: ,
%\end{equation}
%while the interface brane trajectory is determined implicitly by 
%\begin{equation}
%\begin{split}
%    L_{1} f_{1} \frac{dz_{1}^{\textnormal{int}}}{ds} + L_{2} f_{2} \frac{dz_{2}^{\textnormal{int}}}{ds} & = \left( \kappa + \frac{\alpha (d-2)}{2 r^{2}} \left(  \frac{dr}{ds} \right)^{2} \right) r  \: , \qquad L_{i}^{2} f_{i} \left( \frac{dz_{i}^{\textnormal{int}}}{ds} \right)^{2} + \frac{1}{f_{i}} \left( \frac{dr}{ds} \right)^{2} = 1 \: ,
%\end{split}
%\end{equation}
%where $s$ is the proper length along the brane. 

While there may be various constraints on the model parameters, including 
%$\alpha$ and 
$\gamma$, required to ensure that the bulk theory is a reasonable holographic dual of a BCFT, a good starting point is
%may be 
to consider those theories for which the corresponding effective theory enjoys a positive-sign Einstein-Hilbert term. Ideally, one will also have a suppression of the higher curvature terms in the effective theory. 
We should therefore clarify the action for the effective theories describing the physics of the above models. We can do so following the general recipe outlined in \cite{Chen:2020uac}. 

As derived in \cite{deHaro:2000vlm} (see also \cite{Chen:2020uac}), the contribution induced by integrating the bulk action (including the Gibbons-Hawking-York term) on-shell is given by
\begin{equation}
    \begin{split}
        S_{\textnormal{induced}} 
        & = \frac{1}{16 \pi G_{\textnormal{bulk}}} \int d^{d} x \sqrt{-h} \Bigg[ \frac{2 (d-1)}{L_{1}}  + \frac{L_{1}}{(d-2)}  R^{(d)} \\
        & \qquad + \frac{L_{1}^{3}}{(d-4)(d-2)^{2}} \left( R_{ab} R^{ab} - \frac{d}{4(d-1)} R^{2} \right) + \ldots \Bigg] \: .
    \end{split}
\end{equation}
Higher order terms would be expected to depend in detail on the IR physics, including the dynamics of the interface brane.
In fact, we are interested in the case $d=4$, so the last term shown will be modified; we anticipate that the numerical coefficient will be replaced by an order one number, and an additional ``non-local" term of the schematic form ``$R^{2} L_{1}^{3} \ln(RL_{1}^{2})$" will occur. 
The full effective action, including the terms from $S_{\textnormal{ETW}}$, is therefore
\begin{equation}
    \begin{split}
        S_{\textnormal{eff}} 
        & = \frac{1}{16 \pi G_{\textnormal{bulk}}} \int d^{d} x \sqrt{-h} \Bigg[ \frac{2 (d-1)}{L_{1}} \left( 1 - \lambda L_{1} \right) + \frac{L_{1}}{(d-2)} \left( \frac{(d-2) \gamma}{L_{1}} + 1 \right) R^{(d)}  + \ldots \Bigg] \: .
    \end{split}
\end{equation}

Canonically normalizing the Einstein-Hilbert term, we should define an effective Newton constant
\begin{equation} 
    \frac{1}{G_{\textnormal{eff}}} = \frac{1}{G_{\textnormal{bulk}}} \frac{L_{1}}{(d-2)} \left( \frac{(d-2) \gamma}{L_{1}} + 1 \right) \: ,
\end{equation}
obtaining
\begin{equation} \label{eq:normseff}
    \begin{split}
        S_{\textnormal{eff}} 
        & = \frac{1}{16 \pi G_{\textnormal{eff}}} \int d^{d} x \sqrt{-h} \Bigg[ R^{(d)} + \frac{2 (d-1)(d-2)}{L_{1}^{2}} \frac{\left( 1 - \lambda L_{1} \right)}{\frac{(d-2) \gamma}{L_{1}} + 1} + \ldots \Bigg] \: .
    \end{split}
\end{equation}
In particular, the cosmological constant for the effective theory is then
\begin{equation} \label{eq:cc}
    2 \Lambda = - \frac{2 (d-1)(d-2)}{L_{1}^{2}} \frac{(1 - \lambda L_{1})}{(d-2) \frac{\gamma}{L_{1}} + 1} \: ,
\end{equation}
and we must also scale the higher order terms suitably, by replacing $G_{\textnormal{bulk}} \rightarrow G_{\textnormal{eff}} \frac{L_{1}}{(d-2)} \left( \frac{(d-2) \gamma}{L_{1}} + 1 \right)$.

As in the previous section, we would now like to establish the existence of solutions with non-intersecting branes in the NEE limit. 
We begin by considering the special case of a trivial interface, before permitting an interface with non-zero tension.
%$\alpha = 0$. 
%We leave the case $\alpha \neq 0$, the most general interface in this model, to future analysis. 

%One question that we may be interested in is whether the addition of the Einstein-Hilbert term allows us to obtain solutions of the type that we are interested in within a broader region of parameter space. In particular, it might be that even if we choose $\mu < 1$, we can still get the interface brane and the ETW brane to join properly by adjusting the parameter $\alpha$. 

\subsection{Trivial interface}

We will begin by considering the case with only an ETW brane and no interface brane.\footnote{ We are free to drop the subscript on bulk quantities in this subsection, since we have a single region of the AdS soliton.} In this case, we must demand the $z$ coordinate to have the appropriate periodicity $\beta$. While one might hope that the addition of an extra parameter as compared to the model of Section \ref{sec:BHMC} could permit solutions with the property $r_{0}^{\textnormal{ETW}} / r_{\textnormal{H}} \gg 1$, 
%some doubt is cast on this possibility by previously mentioned arguments in the effective theory, where it is anticipated that large quantities of negative energy would be required to source such solutions. We will verify this intuition in the following.
we will see that this does not occur.

We will be interested in the limit where $L \lambda \rightarrow 1$, which we recognize as the critical tension limit where $r_{0}^{\textnormal{ETW}} \rightarrow \infty$ due to (\ref{eq:r0ETW}) (note that the expression for $r_{0}^{\textnormal{ETW}}$ in terms of $\lambda$ is unchanged from the pure tension case); to investigate this limit, we will consider the tension
\begin{equation}
    \lambda L = 1 - \epsilon
\end{equation}
with $\epsilon > 0$ small. At leading order, we find
%\begin{equation}
%    \begin{split}
%        \Delta z & = \frac{2 \pi L_{1}}{ d r_{\textnormal{H}}} \left( \frac{2}{\epsilon}  \right)^{1/2 - 1/d} \mathcal{I}_{0} \sqrt{\frac{(d-2) \gamma}{L} + 1}  %\int_{1}^{\infty} \frac{dy}{y^{1/d}} \frac{1}{\sqrt{y (y-1)}} \: ,
%    \end{split}
%\end{equation}
%taking all parameters other than $\epsilon$ to be fixed, so we find the leading order contribution to the ratio
\begin{equation}
    \frac{2 \Delta z^{\textnormal{ETW}}}{\beta} = \left( \frac{2}{\epsilon}  \right)^{1/2 - 1/d} \mathcal{I}_{0} \sqrt{\frac{(d-2) \gamma}{L} + 1}  \: ,
\end{equation}
taking all parameters other than $\epsilon$ to be fixed. 
To avoid self-intersections, this ratio should be smaller than one; this would appear to be possible provided that we take $\gamma \rightarrow - \frac{L}{(d-2)}$ sufficiently quickly, namely
\begin{equation}
    \left| \frac{(d-2) \gamma}{L} + 1 \right| = O( \epsilon^{1 - 2/d})  \: .
\end{equation}
In particular, we should saturate these asymptotics to avoid sending $\Delta z^{\textnormal{ETW}} / \beta$ to zero.

%Of course, before concluding that this would be sufficient, we must check more carefully, since the leading order contribution to the ratio found above was arrived at assuming that $\gamma$ was fixed. If we take $\gamma = - \frac{L}{(d-2)} \left( 1 - \gamma_{1} \epsilon^{c} \right)$ from the beginning, then we find

Note that, in this case, the cosmological constant for the effective theory (\ref{eq:cc}) will be vanishing in the limit, while our expectation is that the coefficients for the higher curvature terms will blow up, due to the rescaling of coefficients required to obtain the canonically normalized effective action. We are most interested in an effective theory where the higher curvature terms remain under control, so the trivial interface does not appear desirable for our purposes.

\subsection{Non-zero tension interface}

We would now like to consider the case where we restore the interface, but leave the interface brane action as a pure tension term, and take the NEE limit. 
To this end, we again consider near-critical ETW brane tension 
\begin{equation}
    \lambda L_{1} = 1 - \epsilon \: ,
\end{equation}
with $\epsilon > 0$ small. Note that we require (at leading order) $\epsilon < \frac{2e}{1 - \mu u}$ to ensure that the minimum ETW brane radius is larger than that of the interface brane, using the expression (\ref{eq:r0ETW}) for $r_{0}^{\textnormal{ETW}}$ and 
(\ref{eq:asymptotic_r0int}) for $r_{0}^{\textnormal{int}}$ in the NEE limit. 
We then obtain 
\begin{equation}
    \Delta z_{1}^{\textnormal{ETW}} \sim \frac{2 \pi L_{1}}{d r_{\textnormal{H}}^{(1)}} \left( \frac{2}{\epsilon} \right)^{1/2 - 1/d} \mathcal{I}_{0} \sqrt{ \frac{(d-2) \gamma}{L_{1}} + 1 } \: ,
\end{equation}
and thus
\begin{equation}
    \frac{\Delta z_{1}^{\textnormal{ETW}}}{\Delta z_{1}^{\textnormal{int}}} \sim \left( \frac{2}{1 - \mu u} \frac{e}{\epsilon} \right)^{1/2 - 1/d} \sqrt{ \frac{ 1 - \mu u}{1 - u}} \sqrt{ \frac{(d-2) \gamma}{L_{1}} + 1 } \: .
\end{equation}
%In particular, since we require this quantity to be smaller than one, we must have
%\begin{equation}
%    \frac{(d-2) \gamma}{L} <  - \frac{u ( 1  - \mu ) }{ 1 - \mu u} \: .
%\end{equation}
Since we would like to require that this approaches one in the limit, and we have in the limit
\begin{equation}
    \left( \frac{2}{1 - \mu u} \frac{e}{\epsilon} \right)^{1/2 - 1/d} \sqrt{ \frac{ 1 - \mu u}{1 - u}} > 1 \: ,
\end{equation}
we see that this is still a requirement that $\gamma$ be negative in the limit; however, it is less stringent than in the case of a trivial interface. In particular, if we take $\epsilon$ to scale proportionally to $e$ (while keeping $\epsilon < \frac{2e}{1 - \mu u}$ throughout), and recall that $\mu \rightarrow 0$ is required to ensure $R_{2}(u, \mu, e) > 0$, then we see that this bound always requires $\frac{(d-2) \gamma}{L_{1}} + 1$ to approach a positive constant, rather than zero, in the limit. 

Specifically, if we take $\epsilon \sim \frac{2 e c}{1 - \mu u}$ with fixed $0 < c < 1$, then we require%\footnote{Evidently, this implies $-1 < \frac{(d-2) \gamma}{L_{1}} < -u$ in the limit.}
\begin{equation} \label{eq:NEElim_gamma}
    \lim_{\textnormal{NEE}} \left( \frac{(d-2) \gamma}{L_{1}} + 1 \right) = c^{1 - 2/d} \left( 1 - u \right) \: .
\end{equation}
In particular, we see that the limiting value of $\gamma$ lies within the range
\begin{equation} \label{eq:gammarange}
    - 1 < \frac{(d-2) \gamma}{L_{1}} < - u \: .
\end{equation}

The fact that the quantity appearing in (\ref{eq:NEElim_gamma}), which appeared as a scaling factor in the denominator of terms in the properly normalized effective action (\ref{eq:normseff}), is now a positive constant in the limit implies that the coefficients for the higher curvature terms will remain finite. Consequently, for a weakly curved ETW brane, it seems plausible that the physics should be well-described by pure Einstein gravity with small corrections. The cosmological constant for the effective theory again vanishes in the limit. We expect that the curvature length scale of the ETW brane should become parametrically larger than the $(d+1)$-dimensional AdS scale in the limit, with the ratio diverging in the strict limit.

%We find a coupling for the Einstein-Hilbert term in the action for the effective theory which satisfies
%\begin{equation}
%    \frac{1}{16 \pi G_{\textnormal{eff}}} \sim \frac{c^{1 - 2/d}}{16 \pi G_{\textnormal{bulk}}} \frac{L_{1} - L_{2}}{(d-2)}  \: ,
%\end{equation}
%so the effective coupling in the limit is controlled by the positive difference between the central charges of the two CFTs. 

%Meanwhile, the cosmological constant for the effective theory will vanish in the limit, while the coefficients for the higher curvature terms will remain finite. Consequently, for a weakly curved ETW brane, we expect that the physics of the brane should be well-described by pure Einstein gravity. 

Here we have shown that it is possible to indicate a limit for which one can obtain a solution with properly joining branes, for which the minimum radius of the ETW brane is larger than that of the interface brane.
This limit can be interpreted as taking the NEE limit while tuning the ETW brane tension so that the brane propagates close to the asymptotic AdS boundary, and tuning the Einstein-Hilbert or DGP term so that the ETW and interface branes join properly; it is given by
\begin{equation} \label{eq:cosmolimit}
    \boxed{\mu \rightarrow 0 \: , \qquad e \rightarrow 0 \: , \qquad 1 - \lambda L_{1} \sim 2 e c \: , \qquad \left( \frac{(d-2) \gamma}{L_{1}} + 1 \right) \sim c^{1 - 2/d} (1 - u)} \: ,
\end{equation}
where we keep $0 < u < 1$ and $0<c<1$ fixed. Here, one must take $\mu$ to simply vanish sufficiently quickly so that $R_{2}$ remains positive in the limit, meaning that $\mu = O(e^{\frac{d}{2} - 1})$. 
Note that we have yet to establish that the ETW brane stays outside of the interface brane, i.e. that the branes do not intersect, in order to verify that the desired solutions indeed exist. 
We verify this property 
%for a situation where $\mu$ vanishes strictly more quickly than $e^{\frac{d}{2} - 1}$ 
in Appendix \ref{app:intersection}.\footnote{In particular, we verify that it holds for $d \geq 4$, including the case $d=4$ we are especially interested in.} 
%this corresponds to a situation where $x$ is not fixed in the limit, but rather becomes arbitrarily large. We leave the case of fixed $x$ to future work. 

%An important point to emphasize in this analysis is that the solutions we are attempting to construct appear to require negative $\gamma$, and in particular a range
%\begin{equation} \label{eq:gammarange}
%    - 1 < \frac{(d-2) \gamma}{L_{1}} < - u \: .
%\end{equation}
%It has been suggested that such values of this parameter may be problematic for holographic models of this type; for example, it was noted in Appendix B of \cite{Chen:2020uac} that such models may permit the formation of ``Ryu-Takayanagi bubbles" on the brane whose associated generalized entropy may be negative.\footnote{We thank Dominik Neuenfeld for emphasizing this and related points.} 

%We note that this limit appears as a generalization of the NEE limit in the case where we have also introduced an ETW brane in the bulk; here, we must simultaneously tune the ETW brane tension toward criticality, and tune the DGP term to recover a fixed, positive value of the coefficient of the effective Einstein-Hilbert term. 

We note in passing that, for the limit considered here, the coupling for the Einstein-Hilbert term in the action for the effective theory satisfies
\begin{equation}
    \frac{1}{16 \pi G_{\textnormal{eff}}} \sim \frac{c^{1 - 2/d}}{16 \pi G_{\textnormal{bulk}}} \frac{L_{1} - L_{2}}{(d-2)}  \: ,
\end{equation}
so the effective coupling in the limit is controlled by the positive difference between the central charges of the two CFTs.

\section{Conclusions} \label{sec:conclusion}

In this work, we have pursued the suggestion of \cite{VanRaamsdonk:2021qgv} that adding an interface brane to the existing bottom-up holographic models in \cite{Cooper:2018cmb, VanRaamsdonk:2020tlr, VanRaamsdonk:2021qgv} could permit solutions capable of realizing localized gravity on an ETW brane via the Karch/Randall/Sundrum mechanism, making such solutions ``cosmologically viable". We provide evidence to affirm this suggestion, with an important caveat: one also needs to include additional local geometrical terms in the ETW brane action, such as an Einstein-Hilbert term. In particular, just adding a constant tension interface brane (with no Einstein-Hilbert term on the ETW brane) was not sufficient, and just adding an Einstein-Hilbert term to the ETW brane (with no interface brane) was also not sufficient. 

With both ingredients, we found that solutions appear in the region of parameter space, the ``NEE limit", associated with cosmologically viable solutions; this represents an important proof-of-concept for these models. 
Solutions in this limit require a ``wrong sign" Einstein-Hilbert term on the ETW brane, as indicated in (\ref{eq:cosmolimit}) and (\ref{eq:gammarange}), but correspond to a ``correct sign" Einstein-Hilbert term in the action describing the physics of the effective theory. While the latter is the most important criterion for ensuring a physically reasonable model (given that the effective theory is where the cosmology lives), one may still wonder whether there may be other important constraints on the parameters involved in this model arising from the requirement that the bulk physics represents a valid holographic dual of a BCFT. 
Indeed, it has been suggested that such negative values of the ``DGP coupling" parameter may be problematic for holographic models of this type; for example, it was noted in Appendix B of \cite{Chen:2020uac} that such models may permit the formation of ``Ryu-Takayanagi bubbles" on the brane whose associated generalized entropy may be negative, an evident pathology.\footnote{We thank Dominik Neuenfeld for emphasizing this and related points.} 
We leave the interesting question of better understanding these possible additional constraints to future work.

\section*{Acknowledgments}

The author would like to thank Mark Van Raamsdonk for early collaboration and comments on the draft, Dominik Neuenfeld for helpful comments, %and Rob Myers for useful ``discussions-by-proxy".  
and Seamus Fallows and Simon Ross for coordinating submissions on the arXiv. The author is supported by a PGS-D scholarship from the National Sciences and Engineering Research Council of Canada, and by a Four-Year Doctoral Fellowship from the University of British Columbia.

\appendix

\section{Brane trajectories} \label{app:branetrajectories}

Throughout this appendix, we will be interested in a codimension-1 surface parametrized by $(z, r, x^{\mu}) = (Z(r), r, x^{\mu})$ in the AdS soliton geometry
\begin{equation}
    ds_{d+1}^{2} = L^{2} f(r) dz^{2} + \frac{dr^{2}}{f(r)} + r^{2} dx_{\mu} dx^{\mu} \: .
\end{equation}
This may be either an interface brane or an ETW brane; the calculation of intrinsic geometrical quantities and the extrinsic curvature with respect to one side will be identical in both cases, so we will not distinguish between these cases until we come to the equations of motion. We also suppress the coordinate subscripts that would differentiate between the regions $\mathcal{M}_{1}$ and $\mathcal{M}_{2}$ in the interface case. We could allow $dx_{\mu} dx^{\mu} = \eta_{\mu \nu} dx^{\mu} dx^{\nu}$ to denote the metric on either flat Euclidean or Minkowski space; the choice of signature will not affect any of the expressions we derive. 

\subsection*{Geometrical quantities}

We have tangent vector
\begin{equation}
    e_{r}^{\mu} = (Z'(r), 1, \vec{0}) \: , 
\end{equation}
and the rest of the tangent vectors on the brane are just unit vectors spanning the $x^{\mu}$ directions. The induced metric $h_{ab}$ on the ETW brane is of course
\begin{equation}
    ds_{d}^{2} =  \frac{L^{2}}{c(r)^{2}} dr^{2} + r^{2} d x_{\mu} d x^{\mu} \: , \qquad c(r) \equiv \sqrt{\frac{L^{2} f(r)}{1 + L^{2} f(r)^{2} (Z'(r))^{2}}}  \: .
\end{equation}

The spacelike unit normal vector to the brane with the correct orientation (pointing out of the region) is given by
\begin{equation}
    n_{\mu} = c(r) (- 1 , Z'(r), \vec{0}) \: .
\end{equation}
We can now compute the extrinsic curvature
\begin{equation}
    K_{ab} = e_{a}^{\mu} e_{b}^{\nu} \nabla_{\mu} n_{\nu} \: ,
\end{equation}
using that
\begin{equation}
\begin{split}
    \nabla_{\mu} n_{\nu} & = \frac{L^{2} c f f' Z'}{2} dz^{2} + \left( \frac{c f'}{2f} - c' \right) dz \: dr + \frac{c f'}{2f} dr \: dz \\
    & \qquad + \left( \frac{c f' Z'}{2f}  + c' Z' + c Z'' \right) dr^{2} + r c f Z' dx_{\mu} dx^{\mu} \: .
\end{split}
\end{equation}
We find
\begin{equation}
    \begin{split}
        K_{rr} & = e_{r}^{r} e_{r}^{r} \nabla_{r} n_{r} + e_{r}^{z} e_{r}^{r} \nabla_{z} n_{r} + e_{r}^{r} e_{r}^{z} \nabla_{r} n_{z} + e_{r}^{z} e_{r}^{z} \nabla_{z} n_{z} \\
        & = c \left( Z'' + \frac{f' Z'}{2f} (L^{2} f^{2} (Z')^{2} + 3) \right) \\
        K_{ii} & = r c f Z' \eta_{ii} \: ,
    \end{split}
\end{equation}
with all other components vanishing; here, the $i$ appearing in $K_{ii}$ is an (unsummed) $(d-1)$-dimensional Lorentz index. In particular, the scalar extrinsic curvature is
\begin{equation}
    K = h^{ab} K_{ab} = \frac{c^{3}}{L^{2}} \left( Z'' + \frac{f' Z'}{2f} (L^{2} f^{2} (Z')^{2} + 3) \right) + \frac{(d-1)}{r} c f Z' \: .
\end{equation}

In some cases, it may be useful to phrase our analysis in terms of derivatives with respect to a proper length coordinate $s$ along the brane in the $(z, r)$-plane; that is, we take this to be the coordinate appearing in our intrinsic parametrization of the brane, which then has metric
\begin{equation}
    ds_{d}^{2} = ds^{2} + r(s)^{2} dx_{\mu} dx^{\mu} \: .
\end{equation}
Such a coordinate is defined by
\begin{equation}
    L^{2} f(r) \left( \frac{dz}{ds} \right)^{2} + \frac{1}{f(r)}  \left( \frac{dr}{ds} \right)^{2} = 1 \: .
\end{equation}
We then express the normal vector as $n_{\mu} = L (- \dot{r}, \dot{z}, \vec{0})$, so the non-vanishing components of the extrinsic curvature may be written as
\begin{equation}
    \begin{split}
        K_{ss} & = \frac{L}{2} \frac{dz}{ds} f'(r) \left( 3 - L^{2} f(r) \left( \frac{dz}{ds} \right)^{2} \right) \\
        K_{ii} & = L r f(r) \frac{dz}{ds} \eta_{ii} \: .
    \end{split}
\end{equation}

We note that reversing the orientation of the normal vector used in the definition of the extrinsic curvature has the effect of reversing its sign; this is especially important to note when deducing the interface equation of motion. 

We will also be interested in features of the intrinsic geometry of the brane, namely the components of the Ricci tensor and the Ricci scalar. We find non-vanishing components
\begin{equation}
    R_{rr}^{(d)} = - \frac{(d-1)}{r} \frac{c'(r)}{c(r)} \: , \qquad R_{ii}^{(d)} = - \frac{c(r)^{2}}{L^{2}} \left( (d-2) + r \frac{c'(r)}{c(r)} \right) \eta_{ii} \: ,
\end{equation}
or, in the proper length coordinates,
\begin{equation}
    R_{ss}^{(d)} = - (d-1) \frac{r''(s)}{r(s)} \: , \qquad R_{ii} = - r(s)^{2} \left( \frac{r''(s)}{r(s)} + (d-2) \frac{r'(s)^{2}}{r(s)^{2}} \right) \eta_{ii} \: .
\end{equation}
The Ricci scalars are
\begin{equation}
    R^{(d)} = - (d-1) \frac{c(r)^{2}}{r^{2} L^{2}} \left( (d-2) + 2 r \frac{c'(r)}{c(r)} \right) = - (d-1) \left( 2 \frac{r''(s)}{r(s)} + (d-2) \frac{r'(s)^{2}}{r(s)^{2}} \right) \: .
\end{equation}

%\begin{equation}
%    \begin{split}
%        R_{rr}^{(d)} - \frac{1}{2} R^{(d)} h_{rr} & = \frac{ (d-1)(d-2)}{2 r^{2}}  \\
%        R_{ii}^{(d)} - \frac{1}{2} R^{(d)} h_{\mu \mu} & = \frac{(d-2) c(r)^{2}}{L^{2}}  \left( r \frac{c'(r)}{c(r)} + \frac{(d-3)}{2}  \right) \eta_{ii} \: ,
%    \end{split}
%\end{equation}
%or, in the proper length coordinates, 
%\begin{equation}
%    \begin{split}
%        R_{ss}^{(d)} - \frac{1}{2} R^{(d)} h_{ss} & = \frac{(d-1)(d-2)}{2} \frac{r'(s)^{2}}{r(s)^{2}} \\
%        R_{ii}^{(d)} - \frac{1}{2} R^{(d)} h_{ii} & = (d-2) r(s)^{2} \left( \frac{r''(s)}{r(s)} + \frac{(d-3)}{2} \frac{r'(s)^{2}}{r(s)^{2}} \right) \eta_{ii} \: .
%    \end{split}
%\end{equation}

\subsection{Constant tension branes}

We will first consider the case with two branes of constant tension: an interface brane which divides the bulk into regions 1 and 2, and an ETW brane which we add to region 1. 

Suppose we have the Euclidean gravitational action
\begin{equation}
    \begin{split}
        S & = S_{\textnormal{bulk}} + S_{\textnormal{interface}}^{\textnormal{matter}} + S_{\textnormal{ETW}}^{\textnormal{matter}} \\
        S_{\textnormal{bulk}} & = \frac{1}{16 \pi G_{\textnormal{bulk}}} \sum_{i=1}^{2} \int_{\mathcal{M}_{i}} d^{d+1} x \sqrt{g} \: \left( R - 2 \Lambda_{i} \right) \\
        & \qquad + \frac{1}{8 \pi G_{\textnormal{bulk}}} \int_{\textnormal{interface}} d^{d} y \sqrt{h} \: \left[ K \right] + \frac{1}{8 \pi G_{\textnormal{bulk}}} \int_{\textnormal{ETW}} d^{d} y \sqrt{h} \: K \: ,
    \end{split}
\end{equation}
where we take the brane matter actions to be
\begin{equation}
    S_{\textnormal{interface}}^{\textnormal{matter}} = \frac{(1 - d) \kappa}{8 \pi G_{\textnormal{bulk}}} \int_{\textnormal{interface}} d^{d} y \sqrt{ h} \: , \quad S_{\textnormal{ETW}}^{\textnormal{matter}} = \frac{(1-d) \lambda}{8 \pi G_{\textnormal{bulk}}} \int_{\textnormal{ETW}} d^{d} y \sqrt{ h } \: .
\end{equation}
Here and in the following, the brackets represent the discontinuity $[X] = X_{1} - X_{2}$ across the interface brane. We are also permitting two different cosmological constants $\Lambda_{i}$, related to the AdS lengths $L_{i}$ by
\begin{equation}
    \Lambda_{i} = - \frac{d (d-1)}{2 L_{i}} \: .
\end{equation}

The interface brane trajectory is then determined by the junction conditions
\begin{equation}
    \left[ h_{ab} \right] = 0 \: , \quad \left[ K_{ab} - K h_{ab} \right] = 8 \pi G_{\textnormal{bulk}} T_{ab}^{\textnormal{interface}} = (1-d) \kappa h_{ab} \: ,
\end{equation}
where we use
\begin{equation}
    T_{ab}^{\textnormal{interface}} = \frac{2}{\sqrt{h}} \frac{\delta S_{\textnormal{interface}}^{\textnormal{matter}}}{\delta h^{ab}} = \frac{(1-d) \kappa}{8 \pi G_{\textnormal{bulk}}} h_{ab} \: . 
\end{equation}
It can be convenient to rewrite the second junction condition as
\begin{equation}
    \left[ K_{ab} \right] = \kappa h_{ab} \: .
\end{equation}
Meanwhile, the ETW brane trajectory is determined by the equations of motion
\begin{equation}
    K_{ab} - K h_{ab} = 8 \pi G_{\textnormal{N}} T_{ab}^{\textnormal{ETW}}  = (1-d) \lambda h_{ab} \: ,
\end{equation}
where we use
\begin{equation}
    \begin{split}
        T_{ab}^{\textnormal{ETW}} = \frac{2}{\sqrt{h}} \frac{\delta S_{\textnormal{ETW}}^{\textnormal{matter}}}{\delta h^{ab}} = \frac{(1-d) \lambda}{8 \pi G_{\textnormal{bulk}}} h_{ab} \: .
    \end{split}
\end{equation}
We can choose to write this equation as
\begin{equation}
    K_{ab} = \lambda h_{ab} \: .
\end{equation}

Details of the interface solutions can be found in \cite{May:2021xhz}; the upshot is that the first junction condition implies that the $r$ coordinates of the interface brane agree on both sides of the interface, while the second junction condition yields
\begin{equation}
    L_{1} f_{1} \frac{dz_{1}^{\textnormal{int}}}{ds} + L_{2} f_{2} \frac{dz_{2}^{\textnormal{int}}}{ds} = \kappa r \: . 
\end{equation}
Using the relations
\begin{equation} \label{eq:properlengthdef}
    L_{i}^{2} f_{i} \left( \frac{dz_{i}^{\textnormal{int}}}{ds} \right)^{2} + \frac{1}{f_{i}} \left( \frac{dr}{ds} \right)^{2} = 1 \: ,
\end{equation}
we can rephrase this in terms of $r$-derivatives as
\begin{equation}
    \begin{split}
        L_{1} \frac{dz_{1}^{\textnormal{int}}}{dr} & = - \frac{1}{f_{1} \sqrt{V_{\textnormal{eff}}}} \left( \frac{1}{2 \kappa r} (f_{1} - f_{2} ) + \frac{1}{2} \kappa r \right) \\
        L_{2} \frac{dz_{2}^{\textnormal{int}}}{dr} & = \frac{1}{f_{2} \sqrt{V_{\textnormal{eff}}}} \left( \frac{1}{2 \kappa r} (f_{2} - f_{1} ) + \frac{1}{2} \kappa r \right) \: ,
    \end{split}
\end{equation}
where 
\begin{equation}
    V_{\textnormal{eff}}(r) = f_{1} - \left( \frac{f_{2}  - f_{1} - \kappa^{2} r^{2}}{2 \kappa r} \right)^{2} \: .
\end{equation}

For the ETW brane, we obtain the $rr$-component equation of motion
\begin{equation}
    c_{1}(r) f_{1}(r) \frac{dz_{1}^{\textnormal{ETW}}}{dr} = r \lambda \: .
\end{equation}
Isolating $\frac{dz_{1}^{\textnormal{ETW}}}{dr}$, we obtain
\begin{equation}
    \frac{dz_{1}^{\textnormal{ETW}}}{dr}  = \frac{r \lambda}{L_{1} f_{1}(r)} \frac{1}{\sqrt{f_{1}(r) - r^{2} \lambda^{2}}} \: .
\end{equation}
Substituting this into any of the other equations of motion, we verify that these equations are also satisfied. 
These equations are similar to those obtained in the \cite{Cooper:2018cmb}, though here we consider $(d-1)$-dimensional planar rather than spherical symmetry. 

\subsection{Branes with an Einstein-Hilbert term} \label{app:EHeqs}

We would now like to generalize the setup of the previous subsection by introducing Einstein-Hilbert terms on the branes. In particular, we will now modify the brane actions to
\begin{equation}
    \begin{split}
        S_{\textnormal{interface}} & = \frac{1}{16 \pi G_{\textnormal{interface}}} \int_{\textnormal{interface}} d^{d} y \sqrt{h} \: R^{(d)} + S_{\textnormal{interface}}^{\textnormal{matter}} \\
        S_{\textnormal{ETW}} & = \frac{1}{16 \pi G_{\textnormal{ETW}}} \int_{\textnormal{ETW}} d^{d} y \sqrt{h} \: R^{(d)} + S_{\textnormal{ETW}}^{\textnormal{matter}} \: ,
    \end{split}
\end{equation}
where we will introduce the constants $\alpha, \gamma$ defined by
\begin{equation}
    \frac{1}{G_{\textnormal{interface}}} = \frac{\alpha}{G_{\textnormal{bulk}}} \: , \qquad \frac{1}{G_{\textnormal{ETW}}} = \frac{\gamma}{G_{\textnormal{bulk}}} \: .
\end{equation}

The Israel junction conditions at the interface then yield
\begin{equation}
    \left[ h_{ab} \right] = 0 \: , \qquad \left[ K_{ab} - K h_{ab} \right] = 8 \pi G_{\textnormal{bulk}} T_{ab} \: , \quad T_{ab} \equiv \frac{2}{\sqrt{h}} \frac{\delta S_{\textnormal{interface}}}{\delta h^{ab}} \: .
\end{equation}
Notably, this can be interpreted as saying that the junction conditions are unaffected by the presence of the Einstein-Hilbert term on the brane except through the modification of the energy-momentum tensor (see Section 2.4 of \cite{Chen:2020uac}), which is now
\begin{equation}
    T_{ab} = \frac{ (1-d) \kappa}{8 \pi G_{\textnormal{bulk}}} h_{ab} - \frac{1}{8 \pi G_{\textnormal{interface}}} \left( R_{ab}^{(d)} - \frac{1}{2} R^{(d)} h_{ab} \right) \: .
\end{equation}
All together, we have
\begin{equation}
    [K_{ab} ]  = \kappa h_{ab} - \alpha \left( R_{ab}^{(d)} - \frac{1}{2(d-1)} R^{(d)} h_{ab} \right)  \: .
\end{equation}
On the other hand, the equation of motion for the ETW brane is
\begin{equation}
    K_{ab} - K h_{ab}  =  (1-d) \lambda h_{ab} - \gamma \left( R_{ab}^{(d)} - \frac{1}{2} R^{(d)} h_{ab} \right) \: ,
\end{equation}
which we may also write as
\begin{equation}
    K_{ab} = \lambda h_{ab} - \gamma \left( R_{ab}^{(d)} - \frac{1}{2(d-1)} R^{(d)} h_{ab} \right) \: .
\end{equation}

\subsection*{Interface brane}

As in the constant tension case, the first junction condition for the interface brane again implies that the $r$ coordinate of the interface brane agrees on both sides of the interface brane. Now the second junction condition yields, in terms of the proper length parametrization, 
\begin{equation}
\begin{split}
    L_{1} f_{1} \frac{dz_{1}^{\textnormal{int}}}{ds} + L_{2} f_{2} \frac{dz_{2}^{\textnormal{int}}}{ds} & = \left( \kappa + \frac{\alpha (d-2)}{2 r^{2}} \left(  \frac{dr}{ds} \right)^{2} \right) r  \: .
\end{split}
\end{equation}
As before, we can combine this with the expressions (\ref{eq:properlengthdef}) to determine the derivatives of $z_{1}^{\textnormal{int}}, z_{2}^{\textnormal{int}}$ with respect to $r$; we find
\begin{equation}
    \left( \frac{dr}{ds} \right)^{2} = f_{2} - y(r)^{2} \: ,
\end{equation}
where $y(r)$ is a root of the equation
\begin{equation}
\begin{split}
    & \alpha^{2} (d-2)^{2} y^{4} - 4 \alpha (d-2) r y^{3} - 2 (d-2) \alpha \left( \alpha (d-2) f_{2} + 2  \kappa r^{2} \right) y^{2} + 4 r \left( \alpha (d-2) f_{2} + 2 \kappa r^{2} \right) y \\
    & \qquad + \alpha^{2} (d-2)^{2} f_{2}^{2} + 4 \alpha (d-2) f_{2} \kappa r^{2} + 4 \kappa^{2} r^{4} - 4 (f_{1} - f_{2}) r^{2} = 0 \: .
\end{split}
\end{equation}

%we can first write
%\begin{equation}
%    \begin{split}
%        \left( \kappa + \frac{\alpha (d-2)}{2 r^{2}} \left(  \frac{dr}{ds} \right)^{2} \right) r & = \sqrt{f_{1} - r'(s)^{2}} - \sqrt{f_{2} - r'(s)^{2}} \\
%        \left( \left( \kappa + \frac{\alpha (d-2)}{2 r^{2}} \left(  \frac{dr}{ds} \right)^{2} \right)^{2} r^{2} - f_{1} -  f_{2} + 2 r'(s)^{2} \right)^{2} & =  4 \left( f_{1} - r'(s)^{2} \right) \left( f_{2} - r'(s)^{2} \right)
%    \end{split}
%\end{equation}

%Rather than deriving the $r$-derivative expressions from this, it might be most efficient to derive them directly in the alternative coordinate system. The $rr$-component of the equations of motion yields
%\begin{equation}
%    \begin{split}
%        ssssssss
%    \end{split}
%\end{equation}

\subsection*{ETW brane}

For the ETW brane, we find the $ii$-component equation of motion
\begin{equation}
    f_{1}(r) \frac{d z_{1}^{\textnormal{ETW}}}{dr} = \frac{\lambda r}{c_{1}(r)} + \frac{\gamma (d-2)}{2 L_{1}^{2}} \frac{c_{1}(r)}{r} \: .
\end{equation}
and the $rr$-component
\begin{equation}
    \begin{split}
        & (d-2) r c_{1}(r) f_{1}(r) \frac{d z_{1}^{\textnormal{ETW}}}{dr} + \frac{c_{1}(r)^{3}}{L_{1}^{2}} \left( \frac{d^{2} z_{1}^{\textnormal{ETW}}}{dr^{2}} + \frac{f_{1}'(r) \frac{d z_{1}^{\textnormal{ETW}}}{dr}}{2 f_{1}(r)} (L_{1}^{2} f_{1}^{2}(r) \left( \frac{d z_{1}^{\textnormal{ETW}}}{dr} \right)^{2} + 3) \right) r^{2} \\
        & \qquad = (d-1) \lambda r^{2} + \gamma \frac{(d-2)} {c_{1}(r)^{2}}{L_{1}^{2}} \left( \frac{(d-3)}{2} + r \frac{c_{1}'(r)}{c_{1}(r)} \right) \: .
    \end{split}
\end{equation}
Isolating the derivative $\frac{d z_{1}^{\textnormal{ETW}}}{dr}$ in the first equation, we find
\begin{equation}
    \frac{d z_{1}^{\textnormal{ETW}}}{dr} = \frac{\sqrt{(d-2) \gamma \lambda f_{1}(r) + 2 \lambda^{2} r^{2} - f_{1}(r) + \frac{f_{1}(r)}{r} \sqrt{(d-2)^{2} \gamma^{2} f_{1}(r) + \left( 2 (d-2) \gamma \lambda + 1 \right) r^{2} }}}{\sqrt{2} L_{1} f_{1}(r) \sqrt{ f_{1}(r) - r^{2} \lambda^{2}}} \: .
\end{equation}
%It may be worthwhile to check that this quantity automatically satisfies the $rr$-component equation of motion.  

\section{Monotonicity of \texorpdfstring{$\Delta z_{1}^{\textnormal{ETW}}(\lambda)$}{}} \label{app:monotonicity}

We have the derivative
\begin{equation}
    \begin{split}
        \frac{d}{d\lambda} \Delta z_{1}^{\textnormal{ETW}} (\lambda) & = \lim_{\epsilon \rightarrow 0} \frac{d}{d\lambda} \int_{r_{0}(\lambda) + \epsilon}^{\infty} dr \frac{r \lambda}{L f(r)} \frac{1}{\sqrt{f(r) - r^{2} \lambda^{2}}} \\
        & = \lim_{\epsilon \rightarrow 0} \Bigg[ - \frac{d r_{0}(\lambda)}{d\lambda} \Big[ \frac{r\lambda}{L f(r)} \frac{1}{\sqrt{f(r) - r^{2} \lambda^{2}}} \Big]_{r = r_{0}(\lambda) + \epsilon} \\
        & \qquad \qquad + \frac{1}{L} \int_{r_{0}(\lambda) + \epsilon}^{\infty} dr \:  \frac{r}{\left( f(r) - r^{2} \lambda^{2} \right)^{3/2}} \Bigg] \: ,
    \end{split}
\end{equation}
where we have introduced an IR regulator so that the terms in the derivative as per the Leibniz integral rule are finite, and we are dropping the subscripts 1 and 2 for convenience in this appendix (all quantities involve the ETW brane, which propagates in region 1 only). The first term goes as
\begin{equation}
    - \frac{d r_{0}(\lambda)}{d\lambda} \Big[ \frac{r\lambda}{L f(r)} \frac{1}{\sqrt{f(r) - r^{2} \lambda^{2}}} \Big]_{r = r_{0}(\lambda) + \epsilon} = - \frac{2}{d^{3/2}} \frac{L^{2}}{(1 - L^{2} \lambda^{2})^{3/2}}  \frac{1}{\sqrt{ r_{0}(\lambda) \epsilon }} + O(\sqrt{\epsilon}) \: ,
\end{equation}
while the second goes as
\begin{equation}
    \begin{split}
        \frac{1}{L} \int_{r_{0}(\lambda) + \epsilon}^{\infty} dr \:  \frac{r}{\left( f(r) - r^{2} \lambda^{2} \right)^{3/2}} & = \frac{L^{2}}{r_{0}(\lambda) (1 - L^{2} \lambda^{2})^{3/2}} \Big[ \frac{2}{d^{3/2}} \sqrt{\frac{r_{0}(\lambda)}{\epsilon}} - \frac{2 \sqrt{\pi} \Gamma( \frac{1}{d} + 1)}{\Gamma(\frac{1}{d} - \frac{1}{2}) } \Big] \: ,
    \end{split}
\end{equation}
where we use
\begin{equation}
    \int \frac{dy}{y^{2}} \frac{1}{(1 - y^{-d})^{3/2}} = - \frac{1}{y} {}_{2} F_{1} \left( \frac{3}{2}, \frac{1}{d} ; 1 + \frac{1}{d} ;  y^{-d} \right)
\end{equation}
and
\begin{equation}
     {}_{2} F_{1} \left( \frac{3}{2}, \frac{1}{d} ; 1 + \frac{1}{d} ; \left( 1 + \frac{\epsilon}{r_{0}}\right)^{-d} \right) = \frac{2}{d^{3/2}} \sqrt{\frac{r_{0}}{\epsilon}} - \frac{2 \sqrt{\pi} \Gamma( \frac{1}{d} + 1)}{\Gamma(\frac{1}{d} - \frac{1}{2}) } + O(\sqrt{\epsilon}) \: .
\end{equation}
We therefore obtain (for $d > 2$)
\begin{equation}
    \frac{d}{d\lambda} \Delta z_{1}^{\textnormal{ETW}}(\lambda) = - \frac{2 \sqrt{\pi} \Gamma(\frac{1}{d} + 1)}{\Gamma(\frac{1}{d} - \frac{1}{2})} \frac{L^{2}}{r_{0}(\lambda) (1 - L^{2} \lambda^{2})^{3/2}} \: ,
\end{equation}
which is manifestly positive, as desired.

\section{Confirmation of ETW/interface non-intersection} \label{app:intersection}

In general, suppose that we have verified that, for a fixed set of parameters $(L_{1}, \mu_{1}, u, \mu, e)$ and $\lambda$, one has
\begin{equation}
    R_{2}(u, \mu, e) > 0 \qquad \textnormal{and} \qquad r_{0}^{\textnormal{ETW}} > r_{0}^{\textnormal{int}} \qquad \textnormal{and} \qquad \frac{\Delta z_{1}^{\textnormal{ETW}}(\lambda)}{\Delta z_{1}^{\textnormal{int}}} = 1 \: .
\end{equation}
This does not yet constitute a demonstration that the solution is well-behaved, because the ETW and interface branes may intersect at some finite $r_{1}$. We would like to verify that this does not occur for the solutions in the limit identified in Section \ref{sec:EH}. 

In general, to verify that there are no intersections for some set of parameters, it suffices to show that
\begin{equation} \label{eq:deriv_ineq}
    (z_{1}^{\textnormal{ETW}} )' (r_{1}) > (z_{1}^{\textnormal{int}} ) ' (r_{1}) \qquad \textnormal{for all } r_{0}^{\textnormal{ETW}} < r_{1} < \infty \: .
\end{equation}
Indeed, if by contradiction we had that the above inequality held and that $z_{1}^{\textnormal{int}}(\tilde{r}_{1}) = z_{1}^{\textnormal{ETW}}(\tilde{r}_{1}) = \tilde{z}$ at some finite $\tilde{r}_{1} > r_{0}^{\textnormal{ETW}}$, then we would obtain
\begin{equation}
    0 = (\Delta z_{1}^{\textnormal{int}}  - \tilde{z}) - (\Delta z_{1}^{\textnormal{ETW}} - \tilde{z}) = \int_{\tilde{r}_{1}}^{\infty} dr_{1} \: \left( (z_{1}^{\textnormal{int}})'(r_{1}) - (z_{1}^{\textnormal{ETW}})'(r_{1}) \right) < 0 \: ,
\end{equation}
which is absurd.

To show that (\ref{eq:deriv_ineq}) holds, it suffices to show that there is no $r_{1} \in (r_{0}^{\textnormal{ETW}} ,  \infty)$ such that $(z_{1}^{\textnormal{ETW}} )' (r_{1}) = (z_{1}^{\textnormal{int}} ) ' (r_{1})$; the fact that the inequality manifestly holds at $r_{1} = r_{0}^{\textnormal{ETW}}$ (where we are comparing a finite quantity to a formally infinite quantity), together with continuity, then implies that the inequality must hold for all finite $r_{1} > r_{0}^{\textnormal{ETW}}$. 

It is straightforward to find all solutions to the equation $(z_{1}^{\textnormal{ETW}} )' (r_{1}) = (z_{1}^{\textnormal{int}} ) ' (r_{1})$ for the models considered in Section \ref{sec:EH}; letting $y = r_{1}^{d}$, we obtain a quartic equation with non-trivial solutions 
\begin{equation}
\begin{split}
    \frac{y}{\mu_{1} L_{1}^{2}} & = \Bigg[ \pm \left( 1 - (1 - 2 e) u \right) \sqrt{a_{1}} + u^{2} \big( (d-2) \gamma \left( \mu - 2 e (1 - e)(1 + \mu) \right) - ( 1 - \mu) \left( 1 - 2 e \right) L_{1} \big) \\
    & \qquad + u \big( - (d-2) \gamma ( 1 - 2 e ) (1 + \mu) + L_{1} (1 - \mu) \big) + (d-2) \gamma \Bigg] \\
    & \qquad \times \Bigg[ - 4 L_{1} (1 - \lambda L_{1}) (1 - u)^{2} + 8 e (1 - u) \left( L_{1} + (d-2) \gamma - 2 u L_{1} (1 - \lambda L_{1}) \right) \\
    & \qquad - 8 e^{2} \left( (d-2) \gamma - 3 u \left( L_{1} + (d-2) \gamma \right) + u^{2} \left( 3 L_{1} \left( 1 - \frac{2}{3} \lambda L_{1} \right) + (d-2) \gamma \right) \right) \\
    & \qquad - 16 u e^{3} \left( (d-2) \gamma - u \left( L_{1} + (d-2) \gamma \right) \right) -8 (d-2) \gamma u^{2} e^{4} \Bigg]^{-1} \\
    \frac{y}{\mu_{1} L_{1}^{2}} & = \Bigg[ \pm \left( 1 - (1 - 2 e) u \right) \sqrt{a_{2}} + u^{2} \big( (d-2) \gamma \left( \mu - 2 e (1 - e)(1 + \mu) \right) + ( 1 - \mu) \left( 1 - 2 e \right) L_{1} \big) \\
    & \qquad + u \big( - (d-2) \gamma ( 1 - 2 e ) (1 + \mu) - L_{1} (1 - \mu) \big) + (d-2) \gamma \Bigg] \\
    & \qquad \times \Bigg[ 4 L_{1} (1 + \lambda L_{1}) (1 - u)^{2} + 8 e (1 - u) \left( - L_{1} + (d-2) \gamma + 2 u L_{1} (1 + \lambda L_{1}) \right) \\
    & \qquad - 8 e^{2} \left( (d-2) \gamma - 3 u \left( - L_{1} + (d-2) \gamma \right) + u^{2} \left( - 3 L_{1} \left( 1 + \frac{2}{3} \lambda L_{1} \right) + (d-2) \gamma \right) \right) \\
    & \qquad - 16 u e^{3} \left( (d-2) \gamma - u \left( - L_{1} + (d-2) \gamma \right) \right) -8 (d-2) \gamma u^{2} e^{4} \Bigg]^{-1} \: ,
\end{split}
\end{equation}
where 
\begin{equation}
    \begin{split}
        a_{1} & = (d-2)^{2} \gamma^{2} \left( 1 + \mu u^{2} (\mu - 4 e (1-e)) - 2 \mu u (1 - 2 e) \right) \\
        & \qquad + 2 (d-2) (1 - \mu) u L_{1} \gamma \left( 1 - u (1 - e - (1 - \mu) \lambda L_{1} ) \right) + u^{2} (1 - \mu)^{2} L_{1}^{2} \\
        a_{2} & = (d-2)^{2} \gamma^{2} \left( 1 + \mu u^{2} (\mu - 4 e (1-e)) - 2 \mu u (1 - 2 e) \right) \\
        & \qquad - 2 (d-2) (1 - \mu) u L_{1} \gamma \left( 1 - u (1 - e + (1 - \mu) \lambda L_{1} ) \right) + u^{2} (1 - \mu)^{2} L_{1}^{2} \: .
    \end{split}
\end{equation}

We are interested in taking the limit identified in Section \ref{sec:EH}, namely
\begin{equation}
    %\mu \sim \mu_{0} e^{\frac{d}{2} - 1} \: , \qquad 
    1 - \lambda L_{1} = \epsilon \sim \frac{2 e c}{1 - \mu u} \: , \qquad \frac{(d-2) \gamma}{L_{1}} + 1 \sim c^{1 - 2/d} (1 - u) \: .
\end{equation}
We also need to take the limit $\mu \rightarrow 0$ sufficiently quickly, so that $\mu = O(e^{\frac{d}{2} - 1})$. In particular, we focus on the case $d \geq 4$, so that $\mu$ vanishes at least linearly in $e$. 
%We also suppose that $\mu \rightarrow 0$ sufficiently quickly that it can be set to zero from the outset; this would correspond to a limit where $x$ becomes large, which is not unreasonable for our application, and will allow us to perform the analysis here more easily. 

We note that one has in the limit
\begin{equation}
    (d-2) \gamma + u L_{1} \sim (c^{1 - 2/d} - 1) (1 - u) L_{1}  < 0 \: .
\end{equation}
We therefore find that the leading order contributions to the solutions are
\begin{equation}
    \begin{split}
        \frac{y}{\mu_{1} L_{1}^{2}} & = \frac{(d-2) \gamma + u L_{1}}{4 e c L_{1} (1-u) (c^{-2/d} - 1)} \\
        \frac{y}{\mu_{1} L_{1}^{2}} & = - \frac{(d-2) \gamma}{4} \frac{u^{2}}{(1 - u)} \frac{1}{(d-2) \gamma + u L_{1}} \\
        \frac{y}{\mu_{1} L_{1}^{2}} & = \frac{1}{8 (1 - u)} \Bigg[ - \sqrt{\left( (d-2) \frac{\gamma}{L_{1}} - u \right)^{2} + 4 (d-2) \frac{\gamma}{L_{1}} u^{2}} - u + (d-2) \frac{\gamma}{L_{1}} \Bigg] \\
        \frac{y}{\mu_{1} L_{1}^{2}} & = \frac{1}{8 (1 - u)} \Bigg[ \sqrt{\left( (d-2) \frac{\gamma}{L_{1}} - u \right)^{2} + 4 (d-2) \frac{\gamma}{L_{1}} u^{2}} - u + (d-2) \frac{\gamma}{L_{1}} \Bigg] \: .
    \end{split}
\end{equation}
It is straightforward to see that all of these quantities are negative, with the first diverging and the last three converging to finite quantities, so we cannot have any intersections at finite $r_{1}$ in this case.

%\bibliographystyle{JHEP}
%\bibliography{BottomUp}

\begin{thebibliography}{10}

\bibitem{Maldacena:1997re}
J.~M. Maldacena, \emph{{The Large N limit of superconformal field theories and
  supergravity}}, \href{http://dx.doi.org/10.1023/A:1026654312961}{\emph{Adv.
  Theor. Math. Phys.} {\bf 2} (1998) 231--252},
  [\href{https://arxiv.org/abs/hep-th/9711200}{{\tt hep-th/9711200}}].

\bibitem{Strominger:2001pn}
A.~Strominger, \emph{{The dS / CFT correspondence}},
  \href{http://dx.doi.org/10.1088/1126-6708/2001/10/034}{\emph{JHEP} {\bf 10}
  (2001) 034}, [\href{https://arxiv.org/abs/hep-th/0106113}{{\tt
  hep-th/0106113}}].

\bibitem{Banks:2001px}
T.~Banks and W.~Fischler, \emph{{An Holographic cosmology}},
  \href{https://arxiv.org/abs/hep-th/0111142}{{\tt hep-th/0111142}}.

\bibitem{Hertog:2004rz}
T.~Hertog and G.~T. Horowitz, \emph{{Towards a big crunch dual}},
  \href{http://dx.doi.org/10.1088/1126-6708/2004/07/073}{\emph{JHEP} {\bf 07}
  (2004) 073}, [\href{https://arxiv.org/abs/hep-th/0406134}{{\tt
  hep-th/0406134}}].

\bibitem{Alishahiha:2004md}
M.~Alishahiha, A.~Karch, E.~Silverstein and D.~Tong, \emph{{The dS/dS
  correspondence}}, \href{http://dx.doi.org/10.1063/1.1848341}{\emph{AIP Conf.
  Proc.} {\bf 743} (2004) 393--409},
  [\href{https://arxiv.org/abs/hep-th/0407125}{{\tt hep-th/0407125}}].

\bibitem{Freivogel:2005qh}
B.~Freivogel, V.~E. Hubeny, A.~Maloney, R.~C. Myers, M.~Rangamani and
  S.~Shenker, \emph{{Inflation in AdS/CFT}},
  \href{http://dx.doi.org/10.1088/1126-6708/2006/03/007}{\emph{JHEP} {\bf 03}
  (2006) 007}, [\href{https://arxiv.org/abs/hep-th/0510046}{{\tt
  hep-th/0510046}}].

\bibitem{McFadden:2009fg}
P.~McFadden and K.~Skenderis, \emph{{Holography for Cosmology}},
  \href{http://dx.doi.org/10.1103/PhysRevD.81.021301}{\emph{Phys. Rev. D} {\bf
  81} (2010) 021301}, [\href{https://arxiv.org/abs/0907.5542}{{\tt
  0907.5542}}].

\bibitem{Cooper:2018cmb}
S.~Cooper, M.~Rozali, B.~Swingle, M.~Van~Raamsdonk, C.~Waddell and D.~Wakeham,
  \emph{{Black hole microstate cosmology}},
  \href{http://dx.doi.org/10.1007/JHEP07(2019)065}{\emph{JHEP} {\bf 07} (2019)
  065}, [\href{https://arxiv.org/abs/1810.10601}{{\tt 1810.10601}}].

\bibitem{Antonini:2019qkt}
S.~Antonini and B.~Swingle, \emph{{Cosmology at the end of the world}},
  \href{http://dx.doi.org/10.1038/s41567-020-0909-6}{\emph{Nature Phys.} {\bf
  16} (2020) 881--886}, [\href{https://arxiv.org/abs/1907.06667}{{\tt
  1907.06667}}].

\bibitem{VanRaamsdonk:2020tlr}
M.~Van~Raamsdonk, \emph{{Comments on wormholes, ensembles, and cosmology}},
  \href{http://dx.doi.org/10.1007/JHEP12(2021)156}{\emph{JHEP} {\bf 12} (2021)
  156}, [\href{https://arxiv.org/abs/2008.02259}{{\tt 2008.02259}}].

\bibitem{VanRaamsdonk:2021qgv}
M.~Van~Raamsdonk, \emph{{Cosmology from confinement?}},
  \href{https://arxiv.org/abs/2102.05057}{{\tt 2102.05057}}.

\bibitem{Karch:2000gx}
A.~Karch and L.~Randall, \emph{{Open and closed string interpretation of SUSY
  CFT's on branes with boundaries}},
  \href{http://dx.doi.org/10.1088/1126-6708/2001/06/063}{\emph{JHEP} {\bf 06}
  (2001) 063}, [\href{https://arxiv.org/abs/hep-th/0105132}{{\tt
  hep-th/0105132}}].

\bibitem{Takayanagi:2011zk}
T.~Takayanagi, \emph{{Holographic Dual of BCFT}},
  \href{http://dx.doi.org/10.1103/PhysRevLett.107.101602}{\emph{Phys. Rev.
  Lett.} {\bf 107} (2011) 101602}, [\href{https://arxiv.org/abs/1105.5165}{{\tt
  1105.5165}}].

\bibitem{Fujita:2011fp}
M.~Fujita, T.~Takayanagi and E.~Tonni, \emph{{Aspects of AdS/BCFT}},
  \href{http://dx.doi.org/10.1007/JHEP11(2011)043}{\emph{JHEP} {\bf 11} (2011)
  043}, [\href{https://arxiv.org/abs/1108.5152}{{\tt 1108.5152}}].

\bibitem{Randall:1999vf}
L.~Randall and R.~Sundrum, \emph{{An Alternative to compactification}},
  \href{http://dx.doi.org/10.1103/PhysRevLett.83.4690}{\emph{Phys. Rev. Lett.}
  {\bf 83} (1999) 4690--4693},
  [\href{https://arxiv.org/abs/hep-th/9906064}{{\tt hep-th/9906064}}].

\bibitem{Karch:2000ct}
A.~Karch and L.~Randall, \emph{{Locally localized gravity}},
  \href{http://dx.doi.org/10.1088/1126-6708/2001/05/008}{\emph{JHEP} {\bf 05}
  (2001) 008}, [\href{https://arxiv.org/abs/hep-th/0011156}{{\tt
  hep-th/0011156}}].

\bibitem{May:2021xhz}
A.~May, P.~Simidzija and M.~Van~Raamsdonk, \emph{{Negative energy enhancement
  in layered holographic conformal field theories}},
  \href{https://arxiv.org/abs/2103.14046}{{\tt 2103.14046}}.

\bibitem{Freivogel:2019lej}
B.~Freivogel, V.~Godet, E.~Morvan, J.~F. Pedraza and A.~Rotundo, \emph{{Lessons
  on eternal traversable wormholes in AdS}},
  \href{http://dx.doi.org/10.1007/JHEP07(2019)122}{\emph{JHEP} {\bf 07} (2019)
  122}, [\href{https://arxiv.org/abs/1903.05732}{{\tt 1903.05732}}].

\bibitem{Chen:2020uac}
H.~Z. Chen, R.~C. Myers, D.~Neuenfeld, I.~A. Reyes and J.~Sandor,
  \emph{{Quantum Extremal Islands Made Easy, Part I: Entanglement on the
  Brane}}, \href{http://dx.doi.org/10.1007/JHEP10(2020)166}{\emph{JHEP} {\bf
  10} (2020) 166}, [\href{https://arxiv.org/abs/2006.04851}{{\tt 2006.04851}}].

\bibitem{Dvali:2000hr}
G.~R. Dvali, G.~Gabadadze and M.~Porrati, \emph{{4-D gravity on a brane in 5-D
  Minkowski space}},
  \href{http://dx.doi.org/10.1016/S0370-2693(00)00669-9}{\emph{Phys. Lett. B}
  {\bf 485} (2000) 208--214}, [\href{https://arxiv.org/abs/hep-th/0005016}{{\tt
  hep-th/0005016}}].

\bibitem{Fallows:2022ioc}
S.~Fallows and S.~F. Ross, \emph{{Constraints on cosmologies inside black
  holes}},  \href{https://arxiv.org/abs/2203.02523}{{\tt 2203.02523}}.

\bibitem{Simidzija:2020ukv}
P.~Simidzija and M.~Van~Raamsdonk, \emph{{Holo-ween}},
  \href{http://dx.doi.org/10.1007/JHEP12(2020)028}{\emph{JHEP} {\bf 12} (2020)
  028}, [\href{https://arxiv.org/abs/2006.13943}{{\tt 2006.13943}}].

\bibitem{Akal:2020wfl}
I.~Akal, Y.~Kusuki, T.~Takayanagi and Z.~Wei, \emph{{Codimension two holography
  for wedges}},
  \href{http://dx.doi.org/10.1103/PhysRevD.102.126007}{\emph{Phys. Rev. D} {\bf
  102} (2020) 126007}, [\href{https://arxiv.org/abs/2007.06800}{{\tt
  2007.06800}}].

\bibitem{VanRaamsdonk:2021duo}
M.~Van~Raamsdonk and C.~Waddell, \emph{{Finding AdS$^{5}$\texttimes{} S$^{5}$
  in 2+1 dimensional SCFT physics}},
  \href{http://dx.doi.org/10.1007/JHEP11(2021)145}{\emph{JHEP} {\bf 11} (2021)
  145}, [\href{https://arxiv.org/abs/2109.04479}{{\tt 2109.04479}}].

\bibitem{deHaro:2000vlm}
S.~de~Haro, S.~N. Solodukhin and K.~Skenderis, \emph{{Holographic
  reconstruction of space-time and renormalization in the AdS / CFT
  correspondence}},
  \href{http://dx.doi.org/10.1007/s002200100381}{\emph{Commun. Math. Phys.}
  {\bf 217} (2001) 595--622}, [\href{https://arxiv.org/abs/hep-th/0002230}{{\tt
  hep-th/0002230}}].

\end{thebibliography}

\providecommand{\href}[2]{#2}\begingroup\raggedright\endgroup

\end{document}